\documentstyle[amsmath,amssymb,graphicx]{article}

\renewcommand{\arraystretch}{1.3}

\catcode`\@=11
\def\marginnote#1{}

\newcount\hour
\newcount\minute
\newtoks\amorpm
\hour=\time\divide\hour by60
\minute=\time{\multiply\hour by60 \global\advance\minute by-\hour}
\edef\standardtime{{\ifnum\hour<12 \global\amorpm={am}%
        \else\global\amorpm={pm}\advance\hour by-12 \fi
        \ifnum\hour=0 \hour=12 \fi
        \number\hour:\ifnum\minute<10 0\fi\number\minute\the\amorpm}}
\edef\militarytime{\number\hour:\ifnum\minute<10 0\fi\number\minute}

\def\draftlabel#1{{\@bsphack\if@filesw {\let\thepage\relax
      \xdef\@gtempa{\write\@auxout{\string
          \newlabel{#1}{{\@currentlabel}{\thepage}}}}}\@gtempa \if@nobreak
    \ifvmode\nobreak\fi\fi\fi\@esphack} \gdef\@eqnlabel{#1}}
    \def\@eqnlabel{}
\def\@vacuum{}
\def\draftmarginnote#1{\marginpar{\raggedright\scriptsize\tt#1}}

\def\draft{
  \oddsidemargin -.5truein
  \def\@oddfoot{\footnotesize \sl preliminary draft \hfil
    \rm\thepage\hfil\sl\today\quad\militarytime}
  \let\@evenfoot\@oddfoot \overfullrule 3pt
    \let\label=\draftlabel
    \let\marginnote=\draftmarginnote
  \def\@eqnnum{(\theequation)\rlap{\kern\marginparsep\tt\@eqnlabel}%
    \global\let\@eqnlabel\@vacuum}

  }

\makeatletter
\newdimen\normalarrayskip              
\newdimen\minarrayskip                 
\normalarrayskip\baselineskip
\minarrayskip\jot
\newif\ifold             \oldtrue            \def\new{\oldfalse}
\def\arraymode{\ifold\relax\else\displaystyle\fi} 
\def\eqnumphantom{\phantom{(\theequation)}}     
\def\@arrayskip{\ifold\baselineskip\z@\lineskip\z@
     \else
     \baselineskip\minarrayskip\lineskip2\minarrayskip\fi}
\def\@arrayclassz{\ifcase \@lastchclass \@acolampacol \or
\@ampacol \or \or \or \@addamp \or
   \@acolampacol \or \@firstampfalse \@acol \fi
\edef\@preamble{\@preamble
  \ifcase \@chnum
     \hfil$\relax\arraymode\@sharp$\hfil
     \or $\relax\arraymode\@sharp$\hfil
     \or \hfil$\relax\arraymode\@sharp$\fi}}
\def\@array[#1]#2{\setbox\@arstrutbox=\hbox{\vrule
     height\arraystretch \ht\strutbox
     depth\arraystretch \dp\strutbox
     width\z@}\@mkpream{#2}\edef\@preamble{\halign
\noexpand\@halignto
\bgroup \tabskip\z@ \@arstrut \@preamble \tabskip\z@ \cr}%
\let\@startpbox\@@startpbox \let\@endpbox\@@endpbox
  \if #1t\vtop \else \if#1b\vbox \else \vcenter \fi\fi
  \bgroup \let\par\relax
  \let\@sharp##\let\protect\relax
  \@arrayskip\@preamble}
\def\eqnarray{\stepcounter{equation}%
              \let\@currentlabel=\theequation
              \global\@eqnswtrue
              \global\@eqcnt\z@
              \tabskip\@centering
              \let\\=\@eqncr

 \halign to \displaywidth\bgroup
    \eqnumphantom\@eqnsel\hskip\@centering
    $\displaystyle \tabskip\z@ {##}$%
    \global\@eqcnt\@ne \hskip 2\arraycolsep
         $\displaystyle\arraymode{##}$\hfil
    \global\@eqcnt\tw@ \hskip 2\arraycolsep
         $\displaystyle\tabskip\z@{##}$\hfil
         \tabskip\@centering
    &{##}\tabskip\z@\cr}
\newfont{\hr}{msbm10}
\newfont{\ams}{msam10}

\def\be{\begin{eqnarray}}
\def\ee{\end{eqnarray}}

\def\beq{\begin{equation}}
\def\eeq{\end{equation}}
\def\ba{\beq\new\begin{array}{c}}
\def\ea{\end{array}\eeq}
\def\be{\ba}
\def\ee{\ea}

\def\p{\partial}

\def\l[{\phantom.[}

\def\theequation{\arabic{section}.\arabic{equation}}

\hoffset= - 1.0in         

\newdimen\linethick  \linethick=0.4pt
\newdimen\hboxitspace    \hboxitspace=5pt
\newdimen\vboxitspace    \vboxitspace=5pt

\def\fr#1{%
\beq\new
\vcenter{
\hrule height\linethick
          \hbox{\vrule width\linethick
                \kern\hboxitspace
                \vbox{\kern\vboxitspace
                      \hbox{$\begin{array}{c}\displaystyle#1
         \end{array}$}%
                      \kern\vboxitspace}%
                \kern\hboxitspace
                \vrule width\linethick}%
          \hrule height\linethick}%
\eeq}

\hoffset= - 1.0in         
\title{{\bf Matching branches of non-perturbative conformal block
at its singularity divisor} \vspace{.5cm}}
\author{{\bf H.Itoyama}\footnote{ {\small {\it
Department of Mathematics and Physics,
Osaka City University} and {\it Osaka City University Advanced Mathematical Institute (OCAMI), Osaka, Japan}};
itoyama@sci.osaka-cu.ac.jp}, \ {\bf A.Mironov}\footnote{ {\small {\it
Lebedev Physics Institute} and {\it ITEP, Moscow, Russia}};
mironov@itep.ru; mironov@lpi.ru}, \ {\bf A.Morozov}\thanks{{\small
{\it ITEP, Moscow, Russia}}; morozov@itep.ru}}

\begin{document}
\setcounter{footnote}{3}

\setcounter{tocdepth}{3}

\maketitle

\vspace{-6.5cm}

\begin{center}
\hfill FIAN/TD-06/14\\
\hfill ITEP/TH-15/14
\end{center}

\vspace{3.5cm}

\begin{abstract}
Conformal block is a function of many variables,
usually represented as a formal series,
with coefficients which are certain matrix elements
in the chiral (e.g. Virasoro) algebra.
Non-perturbative conformal block is a
multi-valued function, defined globally over the
space of dimensions, with many branches
and, perhaps, additional free parameters,
not seen at the perturbative level.
We discuss additional complications of non-perturbative
description, caused by the fact that all the
best studied examples of conformal blocks lie at the
singularity locus in the moduli space (at divisors
of the coefficients or, simply, at zeroes of the Kac
determinant).
A typical example is the Ashkin-Teller point, where
at least two naive non-perturbative expressions
are provided by elliptic Dotsenko-Fateev integral
and by the celebrated Zamolodchikov formula in
terms of theta-constants, and they are different.
The situation is somewhat similar at the Ising and other
minimal model points.
\end{abstract}

\bigskip

\bigskip

\section{Introduction}

\setcounter{equation}{0}

Conformal blocks are the central objects in $2d$ conformal theories
\cite{CFT}: they are holomorphic constituents of the correlation functions,
the latter being decomposed into bilinear combinations of the conformal blocks
with different internal (intermediate) dimensions.
Another ingredient of the theory are the structure constants,
defining the coefficients in these expansions (for many purposes
it is convenient not to include them into the normalization of conformal blocks,
which is then chosen in some other way, more suitable from the point
of view of complex analysis).
The correlation function can be decomposed in several different ways, and the
corresponding conformal blocks are related by {\it linear} transformations
called modular transforms.
Through the free fermion formalism \cite{JM}, certain generating functions
are interpreted as $\tau$-functions of the conventional integrable systems and the hierarchies
of KP/Toda type, generalization of this formalism to the WZNW model \cite{WZNW,GMMOS}
should provide a description as non-Abelian $\tau$-functions of \cite{GKLMM}.
Long ago the conformal blocks were interpreted as {\it states} in the Hilbert space
of $3d$ Chern-Simons theory \cite{CS}, thus providing an important ingredient
of modern QFT approaches \cite{Wit,TR,Rama,MoSmi,MMMknots} to knot theory \cite{knoth}.
More recently, the AGT relations \cite{AGT,AGT1,AGT2,AGTmamo,MMSh,AGTproof,AGTbasis} provided yet another interpretation
of conformal blocks and their straightforward $q$-deformations
\cite{Awata,HS5d,ItoO} as sums over instantons in respectively
$4d$ and $5d$ Yang-Mills theories with
extended supersymmetry \cite{LMNS,Nek,Pest}. The AGT correspondence proved to be useful in both direction: say, for
using the sums over instantons for analysis of minimal models \cite{AGT2,MM,AB} and for using the Zamolodchikov solution \cite{Zamell} 
for analysis of instanton contributions in Seiberg-Witten theory in the conformal point \cite{MMMel,Pog}.
All these applications to quantum field theories in various dimensions
explain the central role of conformal blocks in theoretical physics
and the need for their thorough investigation. It was moved far enough,
but, unfortunately, unfinished by Al.Zamolodchikov \cite{ZamCB,Zamell,ZamAT}.

In this paper we concentrate on, perhaps, the simplest non-trivial of all conformal
blocks: the 4-point spherical one,
usually defined as a formal series
in the double ratio $x=\frac{(x_2-x_1)(x_3-x_4)}{(x_3-x_1)(x_2-x_4)}$
of the four points on the Riemann sphere,
\be
B(x) = \sum_{k=0}^\infty B_kx^k
\label{Bser}
\ee
The coefficients $B_k$ depend on the four external  dimensions
$\Delta_1,\ldots,\Delta_4$, on one internal dimension $\Delta$
and on the central charge $c$.

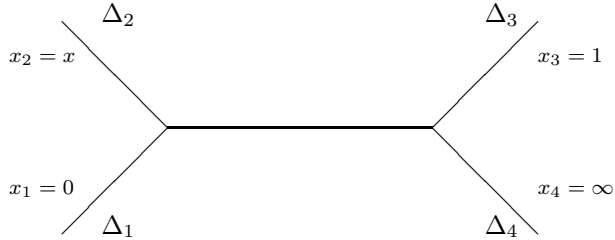
\begin{picture}(200,100)(-150,-50)
\put(0,0){\line(1,0){100}}
\put(-40,40){\line(1,-1){40}}
\put(-40,-40){\line(1,1){40}}
\put(140,40){\line(-1,-1){40}}
\put(140,-40){\line(-1,1){40}}
\put(-25,40){\mbox{$\Delta_2$}}
\put(-25,-40){\mbox{$\Delta_1$}}
\put(120,40){\mbox{$\Delta_3$}}
\put(120,-40){\mbox{$\Delta_4$}}
\put(-60,25){\mbox{{\footnotesize $x_2=x$}}}
\put(-60,-25){\mbox{{\footnotesize $x_1=0$}}}
\put(140,25){\mbox{{\footnotesize $x_3=1$}}}
\put(140,-25){\mbox{{\footnotesize $x_4=\infty$}}}
\end{picture}

\noindent
The series is not everywhere convergent,
thus there is a question of what is the function ${\cal B}(x)$, which has
$B(x)$ as its formal expansion near $x=0$.
We call this function {\it non-perturbative conformal block}
and give a short summary of its already known properties.

As a function of $x$, the 4-point spherical conformal block is believed
to be analytic function ramified just at three points: $0,\ 1,\ \infty$,
with no essential singularities.
Moreover, in the {\it rational} conformal models ramifications are of finite orders,
thus such conformal blocks are actually the Belyi functions \cite{BelGro},
appearing in consideration of Grothendieck's {\it dessins d'enfant}
and the equilateral triangulations.
The ramified coverings of $CP^1$, defined by the Belyi functions, are arithmetic
curves, and their description in terms of conformal models and
rational dimensions is a newly emerging interesting problem
related to description of the universal moduli space \cite{UMS,LevMo}
and to modern theory of the Hurwitz numbers \cite{Hurnu}
and the Hurwitz $\tau$-functions \cite{Hurtau,AMMN}.

Coming back to non-perturbative ${\cal B}(x)$,
the convergence of series (\ref{Bser}) in $x$ is not uniform in the other parameters
$(\Delta_i,\Delta,c)$, and this makes the entire function of all these
variables quite sophisticated.
${\cal B}(x)$ is definitely non-trivial: it has many branches and
changes under modular transformations.
Moreover, this change can be represented as an integral
transform in the internal dimension $\Delta$:
\be\label{mt}
{\cal B}_\Delta(1-x) = \sum_{\Delta'} {\cal M}_\Delta^{\Delta'}{\cal B}_{\Delta'}(x) \\
{\cal B}_\Delta\Big({x\over x-1}\Big) = \sum_{\Delta'} {\cal N}_\Delta^{\Delta'}{\cal B}_{\Delta'}(x)
\ee
and in fact the non-perturbative modular kernels ${\cal M}_\Delta^{\Delta'}$ and
${\cal N}_\Delta^{\Delta'}$ are studied considerably better than ${\cal B}_\Delta(x)$ itself.
The very fact that an $x$-independent modular kernel exists is highly non-trivial.
This happens only if the intermediate (internal) dimensions $\Delta$ is chosen as
a parameter in integral transformation (nothing like this would happen if
we tried to use, say, one of $\Delta_i$ or $c$),
and it reflects the associativity property of
the operator product expansion of conformal field theory (it is also referred to as duality), or of the
co-product in the Virasoro algebra, the modular kernel being the counterpart of the Racah coefficients (or $6j$-symbols)
in the theory of finite-dimensional Lie algebras.

In the context of quantum field theory and, in particular, in CFT
one usually considers a given set of fields, i.e. fixes the set of external dimensions
$\Delta_i$, while $\Delta$ remains arbitrary: it is common in this context
to study the dependence on $\Delta$, but not so common to pay equal attention
at dependencies on $\Delta_i$ or $c$, i.e. at the point in the "space of theories".
Still, in modern stringy approaches it is more than natural just to discuss
the ${\cal B}(\mu|x)$ on entire $6_Cd$-space ${\cal M}$ of parameters $\mu=\{\Delta_i,\Delta,c\}$,
without distinguishing the 1-dimensional "physical slice" ${\cal P}\subset{\cal M}$,
where $\Delta_i,c=const$.
In fact, the conformal block should be analytically continued not only in $x$, but
also in all these six extra parameters $\mu$.
In particular, the modular kernel, which lives on the physical slice
and thus already attracted certain attention,
is believed to possess non-perturbative corrections as
a function of $\Delta$ and $\Delta'$ \cite{Nem2}.

\bigskip

Thus, our goal in this paper is to attract attention at
the non-perturbative conformal block ${\cal B}(\mu|x)$
not only on the physical slice ${\cal P}\subset {\cal M}$, but everywhere else,
with the purpose of identifying the nature and essence
of this important class of special functions.
It turns out that even at the first step in this
direction one runs into interesting details and this
should stimulate more attention to such kind of problems.

\bigskip

There are special choices of external dimensions and central charge,
when much seems to be known about ${\cal B}(x)$, and it is natural
to begin from this point.

(A) For degenerate Verma modules there are null-vector constraints,
which imply differential equations for the conformal block as functions of $x$ \cite{CFT}.
The order of the equation is defined by the level of the null-vector,
and in these cases, it is, first, finite, and, second, greater than one.
The second property means that ${\cal B}(x)$ has different branches
and thus actually lives on a non-trivial Riemann surface in the $x$-space.
The first property means that there is only a {\it finite}-dimensional family
of solutions, while one could think that only one of the six parameters
$\mu$ (that is, a particular external dimension, say, $\Delta_2$)
is fixed by the null-vector condition, and a function of five remaining
parameters should be a solution.

(B) The conformal block possesses the Dotsenko-Fateev representation in terms
of multiple Selberg (generalized hypergeometric) integrals, analytically continued in
the numbers $N_1,\, N_2$ of integration \cite{MMSh}.
However, at $N_1$ and $N_2$ fixed they are just integrals
and thus can be investigated as non-perturbative quantities
by usual means of complex analysis.
This is an obvious possibility, and it was not closely looked at
because the "integral slice" ${\cal I}$ defined by $N_1,N_2=const$
is in a sense transversal to the physical slice ${\cal P}$, and one never looked
carefully at the conformal blocks in transversal directions.

(C) Finally, there is the celebrated Zamolodchikov duality \label{Zam}
applicable to a very special (Ashkin-Teller) model with $c=1$
and $\Delta_i=\frac{1}{16}$.
The point is that the operators of this dimension
create square-root singularities for free fermions
(there are two in the $c=1$ theory),
and thus the correlator of fields $V_{1/16}$ at some points
can be identified with the partition function of free fermions
(Ising model) on a ramified covering of the original space-time.
In the special case of 4-point correlator we get a torus
with $x$-dependent modular parameter $\tau$, and in result
the fantastic formula for the conformal block at Zamolodchikov's slice
${\cal Z}\subset {\cal M}$, with $\Delta_i=\frac{1}{16},\,c=1$:
\be
{\cal B}_\Delta(x) = {\rm Tr}_{\,{\rm free\ fermions}}\ e^{i\pi\tau L_0}
= \frac{q^\Delta}{\theta_{00}(q)}
\ee
where
\be
q=e^{i\pi\tau}, \ \ \ \ x = \frac{\theta_{10}^4(q)}{\theta_{00}^4(q)}
\ee
and $\theta_{00}(q)=\sum_{n=-\infty}^{+\infty} q^{n^2} = \prod_{m=1}^{\infty} (1- q^{2m})(1+ q^{2m-1})^2 $. 
This is again a well-defined non-perturbative formula,
and it is so famous because it is on the physical slice.

\bigskip

In this paper we discuss these examples and interplay between them.
Our main observation is that both the minimal models
(the typical example, where ${\it all}$ external states are null-vectors)
and Zamolodchikov's model at distinguished value of $\Delta=\frac{1}{4}$
appear to lie at the singular locus ${\cal L}\subset{\cal M}$
where coefficients of conformal blocks are ratios of two vanishing functions --
thus they are ambiguously defined.
This is a trivial, but previously underestimated phenomenon,
because it is not seen at the physical slice:
${\cal L}$ and ${\cal P}$ are transversal when intersect.
It is quite interesting to see what happens at these intersections.
Moreover, it is plausible that future investigations of these singularities
will help to understand the non-perturbative dependence of conformal blocks
on $\mu$, which is important in order to shed some light on the somewhat mysterious
non-perturbative expressions for the modular kernels
(postulated by the identification \cite{PT,Zam}
of the  Virasoro Racah matrices with those for peculiar representations
of $SL(2)$).

In order to understand this phenomenon, one has to study the conformal block as a formal power series in $x$
in the vicinities of the singularities.
 They occur at the Kac zeroes, i.e. whenever the special relations between the internal dimension $\Delta$ and the
 central charge $c$ are satisfied. More concretely, if one parameterizes the conformal dimensions with the Dotsenko-Fateev
 parametrization:
 \be
 \Delta=\alpha \Big(\alpha -b+{1\over b}\Big),\ \ \ \ \ \ \ c=1-6\Big(b-{1\over b}\Big)^2
 \ee
 the Kac zeroes in $B_k$ happen at all integer $|m|>0$, $|n|>0$ such that $mn\le k$ at the points
 \be
 \alpha={1\over 2}\Big({n-1)\over b}-(m-1)b)\Big)
 \ee
 In order to have a non-singular answer, one has to require that the conformal
 block has additional zeroes at these singularities. Since the conformal block is a function of $x$, one has to require
 this for each singular $B_k$. In fact, as we shall
 see the Kac zeroes have an embedded structure: once $B_k$ is singular at some $k$, so are
 all $B_{i>k}$. Surprisingly enough, once one imposes the condition of vanishing the numerator of $B_k$ at the Kac zero,
 they also vanish for all $B_{i>k}$! Moreover, despite the answer depends on the direction of approaching the singularity, it
 is parameterized by just one arbitrary constant.

However, in the rational theories a new phenomenon occurs: there accidentally
emerge some higher order poles (higher order Kac zeroes) due to the rational value of the central charge. Again,
when this, say, double pole emerges at some $k$, it is also present at all $B_{i>k}$ until it becomes the third order pole etc.
However, this is generically no longer the case for the corresponding zeroes in the numerator which are needed to cancel these poles:
they are typically simple zeroes even in rational conformal theories, i.e. one meets in these cases the actual singularity.
It means that in a consistent theory the corresponding structure constant should vanish.
This makes the structure of the rational conformal theory in the vicinity of singularity quite sophisticated.

A notable exception is the minimal models. In the case of the minimal models the answer does not depend on the direction
of approaching the singularity! This distinguishes these models and makes them unambiguously defined.

In the paper we illustrate the described picture with the concrete examples of the Ashkin-Teller model and the
minimal models. Note that the Ashkin-Teller model is the only example manifestly known so far, when the answer is ambiguous
near the singularity: the other known examples are the minimal models, when this ambiguity is absent.

\section{Series and "non-perturbative effects"}

\setcounter{equation}{0}

We start with a general discussion of what is the non-perturbative conformal block, i.e. what is the conformal block outside
its divergency radius\footnote{Note here that the divergency radius of the conformal block depends on the chosen variable.
For instance, in variable $x$ this radius does not exceed 1, since there is a singularity at $x=1$. At the same time,
if one uses the variable $q=e^{\pi i\tau}$ related to $x$ by the formula
\be
x = \frac{\theta_{10}^4}{\theta_{00}^4}\\
\theta_{10}(q) = \theta_{10}(0|\tau) = 2q^{1/4}\sum_{n=0} q^{n(n+1)}
= 2q^{1/4} \prod_{m=1}^{\infty} (1- q^{2m})(1 + q^{2m})^2
\ee
the series converge in the entire region where ${\cal B}(x)$ is analytic, 
as was demonstrated by Al.Zamolodchkov \cite{Zamell}.}.
The archetypical example of asymptotic perturbative series
\be
F(x)=\sum_{n=0}^\infty n!\,x^n
\label{facser}
\ee
The factorial growth here is usually connected to the fact that the perturbation
is performed with irrelevant operators,
and a non-perturbative answer should have a different asymptotics at large values of fields.

The most obvious way to handle the series (\ref{facser}) is to interpret the factorials
as the $\Gamma$-function and use its integral representation to define the
non-perturbative function:
\be
{\cal F}(x) = \sum_{n=0}^\infty \Gamma(n+1)x^n = \int_0^\infty \left(\sum_{n=0}^\infty (zx)^n\right)
e^{-z} dz = \int_0^\infty \frac{e^{-z}dz}{1-zx}
\ee
This answer, however, depends on the choice of the contour connecting $0$ and $\infty$,
and the ambiguity is just a residue at the singularity of the integrand:
\be
{\cal F}(x) \ \ \ \ {\rm is\ defined\ modulo} \ \ \ \
\oint_{z=x^{-1}} \frac{e^{-z}dz}{1-zx} \ \sim\ \frac{e^{-1/x}}{x}
\ee

An alternative way to define the non-perturbative function ${\cal F}(x)$ is through
writing an equation satisfied by the perturbative series.
In  the case of (\ref{facser}) the simplest one is the first order  differential equation:
\be
x\frac{d}{dx}\, xF(x) = \sum_{n=0}^\infty (n+1)!\,x^{n+1} = F(x)-1
\ee
Solutions to this equation depend on one free constant: the freedom is
to add (with an arbitrary coefficient) a solution to the homogeneous equation
\be
x^2\frac{dF_0}{dx} = (1-x)F_0(x) \ \ \Longrightarrow \ \ \
F_0(x) = \ \frac{e^{-1/x}}{x}
\ee
i.e. ${\cal F}(x)$ is once again defined modulo $x^{-1}e^{-1/x}$.
The slight difference is that in the first approach the coefficient seems
to be arbitrary integer, while in the second approach, i.e. when
the non-perturbative function is defined as a $D$-module, as a solution to
some linear equation, it can naturally be arbitrary complex number.

Of course, in both cases we considered a {\it minimal} definition:
the non-perturbative function can be lifted to bigger moduli spaces,
for example, by considering equations of higher order, which have more
solutions.
Still, usually there is no a {\it natural} way to get rid of extra parameters;
or, whenever this is done, there should be some {\it additional} reason
for such a restriction. Thus, the moral at this stage is:

\be
{\bf the\ non-perturbative\ function\ usually\ depends\ on\ extra\ {\it hidden}\   parameters}
\ee

\bigskip

We emphasize that this fact is implicit in {\it any} non-perturbative consideration.
It is enough to remind the celebrated example of instantons in Yang-Mills theories:
not only they come with the typical non-perturbative weight $e^{-1/g^2}$,
they bring in an additional parameter, not seen at the perturbative level:
the $\theta$-angle.
In fact, restriction to real angles (i.e. to unimodular {\it free} coefficients
in front of the instanton contribution) is somewhat similar to restriction to
integer-valued coefficients in the above example, i.e. is motivated more by
the concrete method of analytical continuation than by the essence of the
problem.
What is important in this example, it emphasizes the physical relevance of
additional (perturbatively hidden) parameters:
they affect the non-perturbative renormalization \cite{KniM},
whose significance is nowadays well appreciated and widely investigated
with the help of Seiberg-Witten theory \cite{SW1}-\cite{SW3}.

In application to conformal blocks, this
{\it would} imply
that ${\cal B}(\mu|x)$
is actually  ${\cal B}(\mu|x|C)$, i.e. the non-perturbative conformal block
depends on additional parameters $C$
not seen at perturbative level, and not present in the original expansion
(\ref{Bser}) derived from representation theory of the Virasoro algebra.

However, one should expect that the conformal theory is essentially free \cite{DF,FF,W,GMMOS},
therefore there is no room for a factorial growth of expansions in the
coupling constants of the {\it irrelevant} operators. Thus, non-perturbative
hidden parameters do not appear in this context,
and all ambiguities are of a different nature.
As we shall see, this makes the
non-perturbative conformal blocks easier comprehensible than non-perturbative
effects in more general interacting models of quantum field theory.
As already mentioned, the expansion series (\ref{Bser}) in $x$
are believed to have finite radia of convergence,
and in this sense they are somewhat different from (\ref{facser}).

In fact, even for (\ref{facser}) it is recently shown \cite{AMMN}) that it can be
naturally associated with a certain KP/Toda $\tau$-function
(arising in a simplified counting problem of the Belyi functions \cite{Zog}).
Since the $\tau$-functions satisfy {\it quadratic} Hirota relations \cite{Hir,JM,GKLMM}
they are not preserved by {\it linear} transformations, and this can provide
a new interesting tool to restrict non-perturbative ambiguities
(perhaps, even to distinguish the choice of $\theta=0$ in instanton calculus,
thus leading to a new kind of $\theta$-problem solution).

\section{Perturbative expansion of conformal block}

\setcounter{equation}{0}

We begin from reminding the basic facts about the perturbative
conformal block (\ref{Bser}).

\subsection{Definition of  perturbative block for {\it four} primaries}

The basic definition of the 4-point conformal block comes from the bilinear expansion of the correlation function 
of four primary fields \cite{CFT}
\be\label{CBC}
\Big<V_{\Delta_1,\bar\Delta_1}(x_1,\bar x_1)V_{\Delta_2,\bar\Delta_2}(x_2,\bar x_2)V_{\Delta_3,\bar\Delta_3}(x_3,\bar x_3)
V_{\Delta_4,\bar\Delta_4}(x_4,\bar x_4)\Big>=\\
=\sum_{\Delta,\bar\Delta}C^{\Delta,\bar\Delta}_{12}C^{\Delta,\bar\Delta}_{34}\times
G_\Delta(\Delta_1,\Delta_2,\Delta_3,\Delta_4;c;x_1,x_2,x_3,x_4)\times
\bar G_{\bar \Delta}(\bar\Delta_1,\bar\Delta_2,\bar\Delta_3,\bar\Delta_4;c;\bar x_1,\bar x_2,\bar x_3,\bar x_4)
\ee
where $C_{12(34)}^{\Delta,\bar\Delta}$ are the structure constants of the theory which are sometimes included in the definition of the
conformal block. However, in principle, the structure constants is a separate object, it defines the concrete conformal theory,
and it is a separate problem to list all admissible structure constants (they should satisfy additional complicated restrictions like
duality). At the same time, the conformal block $G_\Delta(\Delta_1,\Delta_2,\Delta_3,\Delta_4;c;x_1,x_2,x_3,x_4)$ is a
universal function of four points $x_i$ that depends on 6 parameters: 5 dimensions and the central charge, it encodes only properties
of the Virasoro algebra. In fact, it is a non-trivial function of the double-ratio $x=\frac{(x_2-x_1)(x_3-x_4)}{(x_3-x_1)(x_2-x_4)}$ 
only:
\be
G_\Delta(\Delta_1,\Delta_2,\Delta_3,\Delta_4;c;x_1,x_2,x_3,x_4)
=\Big(\prod_{i<j}x_{ij}^{\zeta_{ij}}\Big)B_\Delta(x)
= \Big(\prod_{i<j}x_{ij}^{\zeta_{ij}}\Big)\sum_k x^kB_k(\Delta_1,\ldots,\Delta_4|\Delta,c)
\ee
Here $x_{ij}\equiv x_i-x_j$, 
$\zeta_{12(13)}=0$, $\zeta_{14}=-2\Delta_1$, $\zeta_{23}=\Delta_4\Delta_1-\Delta_2-\Delta_3$, $\zeta_{24}=\Delta_1+\Delta_3-
\Delta_2-\Delta_4$, $\zeta_{34}=\Delta_1+\Delta_2-\Delta_3-\Delta_4$.
Note that permutations of points gives rise to the modular transformations (\ref{mt}). For instance, upon permuting $x_1$ and $x_3$
we obtain
\be\label{dmt}
B(\Delta_3,\Delta_2,\Delta_1,\Delta_4|\Delta,c;1-x)= \sum_{\Delta'} {\cal M}_\Delta^{\Delta'}
B(\Delta_1,\Delta_2,\Delta_3,\Delta_4|\Delta',c;x)
\ee
From now on we omit the parameters $\Delta_1,\Delta_2,\Delta_3,\Delta_4,c$ in notation of the conformal block unless it may lead
to a misunderstanding.

The conformal block expansion (\ref{CBC}) is derived by using the operator product expansion (OPE)
\be
V_{\Delta_1\Delta_2}(x_1,\bar x_1)V_{\Delta_2,\bar\Delta_2}(x_2,\bar x_2) =
\sum_\Delta C_{12}^{\Delta,\bar\Delta} (x_1-x_2)^{\Delta-\Delta_1-\Delta_2}(\bar x_1-\bar x_2)^{\bar\Delta-\bar\Delta_1-\bar\Delta_2}
\times\\ \times
\left(\sum_{Y,\bar Y}(x_1-x_2)^{|Y|} (\bar x_1-\bar x_2)^{|\bar Y|}\beta_{Y}(\Delta|\Delta_1,\Delta_2)
\bar\beta_{\bar Y}(\bar\Delta|\bar\Delta_1,\bar\Delta_2)
\hat L_{-Y}{\hat {\bar L}}_{-\bar Y} V_{\Delta_2,\bar\Delta_2}(x_2,\bar x_2)\right)
\ee
where the bracket at the r.h.s. is a sum over descendants, which
are labeled by Young diagrams $(Y,\bar Y)$ of all sizes $(|Y|,|\bar Y|$.
For simplicity we consider only the OPE of the primary fields $V_1$ and $V_2$: this is possible
if we restrict ourselves to the 4-point conformal blocks only, which we really do in this paper.

With the OPE, one can project the products $V_{1}(x_1)V_{2}(x_2)$ and $V_{3}(x_3)V_{4}(x_4)$
onto Verma modules given by $V_{\Delta,\bar\Delta}$ and $V_{\Delta',\bar\Delta'}$ correspondingly, 
and since the pair correlation function of
any fields is non-zero iff the both fields have the same conformal dimension, one finally obtains (\ref{CBC}), and the conformal
block represents a chiral part of the correlation function with specified intermediate dimension. It can be
defined completely within the chiral algebra \cite{MS}.

In terms of the chiral algebra one defines the conformal block through a chiral correlator:
\be
B_\Delta(x)=\Big<V_{\Delta_1}(x_1)V_{\Delta_2}(x_2)\Big|_\Delta V_{\Delta}(x_3)V_{\Delta(x_4)}\Big>
\ee
where hereafter we denote the corresponding chiral primary fields with the same letter $V$,
and the chiral OPE looks like
\be\label{cOPE}
V_{\Delta_1}(x_1)V_{\Delta_2}(x_2) \stackrel{\Delta}{\longrightarrow}
 (x_1-x_2)^{\Delta-\Delta_1-\Delta_2}
\left(\sum_{Y} (x_1-x_2)^{|Y|}\beta_Y(\Delta|\Delta_1,\Delta_2)
\hat L_{-Y} V_\Delta(x_1)\right)
\ee
The coefficients $\beta_Y$ are easily related with the three point functions: one suffices to consider the three
point function $\Gamma_Y(\Delta_1,\Delta_2|\Delta)=\Big<V_{\Delta_1}V_{\Delta_2}|\hat L_{-Y}V_\Delta\Big>$ and use (\ref{cOPE}). Then,
one immediately obtains for the three point function
\be
\Gamma_Y(\Delta_1,\Delta_2|\Delta)=\sum_{Y'}T_{Y,Y'}(\Delta)\beta_{Y'}(\Delta|\Delta_1,\Delta_2)
\ee
with the Shapovalov matrix
\be\label{Sh}
T_{Y,Y'}(\Delta)=\Big<\hat L_{-Y} V_\Delta|\hat L_{-Y'}V_\Delta\Big>
\ee

Note that picking up a single term in the sum over $\Delta$ may lead to losing associativity of the product,
which in simplest examples is guaranteed by the structure constants
$C_{12}^{\Delta,\bar\Delta}$. For many models (when there are degenerate representations of the Virasoro algebra in the spectrum)
they vanish for most values of $\Delta,\bar\Delta$. 
However, in these cases it is often enough to just impose selection rules on $\Delta$.
In any case, for the 4-point conformal blocks one can avoid using associativity.
At this stage one gets
\be
B_\Delta(\Delta_i|x) = \sum_{Y,Y'} x_{12}^{|Y|}x_{34}^{|Y'|}
\beta_Y(\Delta|\Delta_1,\Delta_2)\beta_{Y'}(\Delta|\Delta_3,\Delta_4) 
\Big<\hat L_{-Y} V_\Delta \Big|\hat L_{-Y'}V_{\Delta}\Big>=\sum_{Y,Y'}\Gamma_Y(\Delta_1,\Delta_2|\Delta)T_{Y,Y'}^{-1}
\Gamma_{Y'}(\Delta_3,\Delta_4|\Delta)
\label{cbviatwo}
\ee

At last, in order to calculate the 3-point functions, one needs to use properties of the chiral algebra:
\begin{itemize}
\item
consistency of the Virasoro algebra with the scalar product $<...>$:
\be
\Big<\hat L_Y V\Big|V'\Big> \ =\ \Big<V \Big| \hat L_{-Y} V'\Big> 
\ee
for arbitrary operators $V$ and $V'$ (not necessarily primary). We normalize the primaries so that 
$\ \Big<V_\Delta\Big|V_\Delta\Big>\ =1$.
In fact, just this identity implies that the scalar product in (\ref{Sh})
is equal to the Shapovalov matrix
\be
\Big<\hat L_{-Y} V_\Delta \Big|\hat L_{-Y'}V_{\Delta}\Big> =
\Big<V_\Delta\Big| \hat L_{Y}\hat L_{-Y'} V_{\Delta}\Big>,
\ee
in particular it vanishes for Young diagrams of differing sizes $|Y|\neq|\bar Y|$,
what is important for the projective invariance of the correlator, i.e. for
collecting the four coordinates $x_1,\ldots, x_4$ into a single double ratio
$x = \frac{(x_2-x_1)(x_3-x_4)}{(x_3-x_1)(x_2-x_4)}$,
which allows one to put $x_1=0$, $x_2=x$, $x_3=1$, $x_4=\infty$.
\item
The comultiplication for the Virasoro algebra \cite{MS}, which is a direct consequence of the Ward identities \cite{CFT}:
\be
\hat L_n \Big(V_{1}(0) V_2(x)\Big)  = \left(\sum\limits_{k=0}^{\infty}x^{n+1-k}\left(n+1 \atop k\right)\hat L_{k-1}V_1(0)\right)V_2(x) 
 + V_1(0)\hat L_nV_2(x)
\ee
\end{itemize}

Applying these postulates, one immediately obtains \cite{CFT,MMMcb,MMMM}
that for $Y=\{y_1\geq y_2\geq \ldots\}$
\be
\Gamma_Y(\Delta_1,\Delta_2|\Delta) = \prod_i \Big( \Delta+y_i\Delta_2-\Delta_1-\sum_{j<i}y_j\Big)
\ee

\subsection{Chain-vectors}

Alternatively, one can use
a very effective representation of  $B_k$ in (\ref{Bser})
via scalar products of constituents of the distinguished {\it chain-vectors} \cite{CFT}
\be
B_k = \ \Big< k,\Delta_3,\Delta_4\ \Big|\ k,\Delta_1,\Delta_2\Big>
\label{Bsca}
\ee
which depend also on $\Delta$ and $c$ and satisfy simple defining recurrence relations
\be
\hat L_n \,\Big|\ k,\Delta_1,\Delta_2\Big> = \Big(k-n + \Delta + n\Delta_2-\Delta_1\Big)\Big|\ k-n,\Delta_1,\Delta_2\Big>,
\ \ \ \ 0<n\leq k,  \\
\hat L_0 \,\Big|\ k,\Delta_1,\Delta_2\Big> = \Big(k + \Delta \Big)\Big|\ k,\Delta_1,\Delta_2\Big>
\ee
The possibility for such recurrence relations to unambiguously define a chain of such states
within Verma module is a prominent feature of Virasoro algebra, which holds also for
$\widehat{U(1)}$ but not for higher $W_N$ algebras, beginning from $W_3$.
In fact, the chain vector is basically nothing but the projection of the operator product expansion of two primaries with conformal
dimensions $\Delta_1$, $\Delta_2$ onto a third one with dimension $\Delta$:
\be\label{cv}
\hat V_{\Delta_1}(0)\, \hat V_{\Delta_2}(x)\ {\stackrel{\Delta}{\longrightarrow}}\ x^{\Delta-\Delta_1-\Delta_2}\sum_{Y}
x^{|Y|} \beta_Y(\Delta|\Delta_1,\Delta_2) \hat L_{-Y} \hat V_\Delta(0),  \\
\Big|\ k,\Delta_1,\Delta_2\Big> = \sum_{|Y|=k} \beta_Y(\Delta|\Delta_1,\Delta_2) \hat L_{-Y} \hat V_\Delta(0)
\ee
i.e.
\be
B_k= \sum_{|Y_1|=|Y_2|=k} \beta_Y(\Delta|\Delta_1,\Delta_2)\beta_Y(\Delta|\Delta_3,\Delta_4)
\Big< \hat V_{\Delta}\hat L_{Y_1}|\hat L_{-Y_2} \hat V_\Delta(0)\Big>
\ee

Importance of the chain vectors becomes especially clear within the AGT conjecture, since
projectors of the chain vectors for product of the Virasoro and Heisenberg algebras
on peculiar states in the Verma module associated with the generalized Jack polynomials
reproduce the Nekrasov functions,
\be
N_{Y_1,Y_2} =  \Big< k,\Delta_3,\Delta_4\ \Big|\ J_{Y_1,Y_2}\Big>\,\Big<J_{Y_1,Y_2}\ \Big|\ k,\Delta_1,\Delta_2\Big>
\ee
(here pairs of Young diagrams emerge due to the product of the Virasoro and Heisenberg algebras)
and they have extra poles not present in the scalar products (\ref{Bsca}).
For various realizations of this idea see \cite{AGTbasis}.\footnote{For
instance, at the level one:
\be
\Big|\,J_{0,1}\Big>\ =\Big(\hat L_{-1}+(Q+2a)\hat a_{-1}\Big)\Big|\,a\Big>\ \ \ \ \ \ \ \
\Big|\,J_{1,0}\Big>\ =\Big(\hat L_{-1}+(Q-2a)\hat a_{-1}\Big)\Big|\,a\Big>
\ee
i.e.
\be
\hat L_{-1}\Big|\,a\Big>\ = -\frac{1}{4a}
\left(\ (Q-2a)\Big|J_{0,1}\,\Big>-(Q+2a)\Big|J_{1,0}\,\Big>\ \right)\ \ \ \ \ \ \ \
\hat a_{-1}\Big|\,a\Big>\  =\frac{1}{4a}\left(\ \Big|J_{0,1}\,\Big>\ -\ \Big|J_{1,0}\,\Big>\ \right)
\ee
}

\subsection{Explicit expressions}

In result we get for the first several coefficients $B_k$ in
\be
B(x) = 1+\sum_{k=1}^\infty B_k x^k
\label{cbe}
\ee
the following explicit expressions:
\be
B_1 = \frac{(\Delta+\Delta_2-\Delta_1)(\Delta+\Delta_3-\Delta_4)}{2\Delta}
\label{B1}
\ee
\be
{\cal B}_\Delta^{(2)}=
{(\Delta+\Delta_2 -\Delta_1   )(\Delta+\Delta_2 -\Delta_1   +1)
(\Delta+\Delta_3-\Delta_4)
(\Delta+\Delta_3-\Delta_4+1)\over 4\Delta(2\Delta+1)}+\\
+{\left[(\Delta_1   +\Delta_2 )(2\Delta+1)+\Delta(\Delta-1)
-3(\Delta_2   -\Delta_1 )^2\right]
\left[(\Delta_3+\Delta_4)(2\Delta+1)+\Delta(\Delta-1)
-3(\Delta_3-\Delta_4)^2\right]
\over 2(2\Delta+1)\Big(2\Delta(8\Delta-5) + (2\Delta+1)c\Big)}
\label{B2}\\
\ldots
\ee
Denominators of these expressions,
\be
K_2 = 4\Delta(16\Delta^2-10\Delta+2c\Delta+c), \\
K_3=
6(3\Delta^2+c\Delta-7\Delta+c+2)\cdot K_2,  \\
K_4 = 4(8\Delta+c-1)
(16\Delta^2-82\Delta+10c\Delta+15c+66) \cdot K_3,  \\
\ldots
\label{Kred}
\ee
at generic values of $c$ possess only simple zeroes, which coincide with zeroes of the
Kac determinants $KD_n$ (i.e. determinants of Shapovalov matrices), though these latter sometimes have non-zero multiplicities.
These extra (multiplicity of) zeroes of the Kac determinants cancel
against zeroes of the numerators
in expressions for the 4-point conformal blocks, e.g.
\be
{KD}_3=2\Delta\cdot K_3,  \\
{KD}_4=4\Delta^2(16\Delta^2-10\Delta+2c\Delta+c)\cdot K_4
\ee
Since the cancelation takes place at  arbitrary points of the moduli
space ${\cal M}$ (i.e. for arbitrary dimensions and central charges), these zeroes play no role in further our considerations,
and we sometimes call the reduced quantities (\ref{Kred}) the Kac determinants assuming that this
should not cause any confusion.

\subsection{Coefficients $B_k$ from Dotsenko-Fateev representation of \cite{MMSh}}

Let us make the change of variables
\be
\Delta_\mu = \alpha_\mu(\alpha_\mu-Q), c=1-6Q^2, Q = b-\frac{1}{b}
\label{divialpha}
\ee
with $\alpha_i$ constrained by
\be
\alpha-\alpha_1-\alpha_2= b N_1, \\
Q-\alpha -\alpha_3-\alpha_4= bN_2
\label{Nvialpha}
\ee
In fact, one can choose in (\ref{Nvialpha}) $Q-\alpha_i$ instead of any $\alpha_i$, since there is a symmetry in the theory w.r.t.
this operation (in particular, (\ref{divialpha}) remains unchanged under this transformation).

With this change of variables (\ref{divialpha}) $B_k(\Delta_1,\ldots\Delta_4,\Delta,c)$ turns into rational functions
of $J_k(\alpha_1,\alpha_2+\alpha_3,b,N_1,N_2)$,
\be
B_k=J_k
\ee
which at integer non-negative values of $N_1$ and $N_2$ coincide with the
values of Selberg-Kadell\cite{Kad} integrals, $N_1$ times between $0$ and $x$
and $N_2$ times between $0$ and $1$.
This fact \cite{MMSh,ItoT} can be interpreted as Dotsenko-Fateev like representation
of conformal blocks \cite{DF} via conformal matrix model of \cite{conmod,AGTmamo}.

In more detail,
\be
J_k={Z(v)\over Z(0)},\\
Z(v)=\int \prod_{a<a'} (v_a-v_{a'})^{2b^2}
\prod_a v_a^{2\alpha_1b}(x-v_a)^{2\alpha_2 b}(1-v_a)^{2\alpha_3b}
 dv_a
\label{DFint}
\ee
and $N_1$ integrations here runs $0$ to $x$, while $N_2$ integrations goes from $0$ to $1$.
Note that the integrals are not obligatory around the {\it closed} contours:
for irrational products $\alpha_i\alpha_j$ these are not so easy to define.
In other words, following \cite{MMSh} we define the integrals in the same way as
the archetypical
$B$-function integral
$$
\int_0^1 z^{a-1} (1-z)^{b-1} dz = \frac{\Gamma(a)\Gamma(b)}{\Gamma(a+b)}
$$
is defined.
The new point is analytical continuation in $N_1$ and $N_2$:
since $B_k$ are {\it rational} functions of these variables, continuation
is straightforward and unambiguous.

These formulas can also be straightforwardly $q$-deformed \cite{HS5d,ItoO},
this generalization is related to $5d$ gauge theories, to $q$-Virasoro algebras,
to MacDonald polynomials and DAHA.

\subsection{General structure of $B_k$}

The poles of $B_k$ (Kac zeroes) have a very simple form in the Dotsenko-Fateev parametrization: they occur at
all integers $|m|>0$, $|n|>0$ such that $mn\le k$ at the points
 \be\label{Kz}
 \alpha_{m,n}={1\over 2}\Big({n-1\over b}-(m-1)b\Big)
 \ee
It follows that once a pole appears at some $B_k$, it also happens at all $B_{l>k}$. Such a singularity in the conformal block at
level $k$ can be removed by a proper choice of the external dimensions $\Delta_i$ so that the theory would still make sense.
Surprisingly enough, once the pole disappears in this way at the level $k$, it simultaneously disappears at all higher levels.
In other words, not only the Kac zeroes, but also the zeroes of the numerators have a nested structure. In the next sections we
study this phenomenon is detail, in particular, we observe that in conformal theories with rational values of the central charge
there emerge multiple Kac zeroes, which sometimes still cancel out with multiple zeroes of the numerator, but sometimes not.

Our goal in the next sections is to look for a non-perturbative parameters in the conformal block, i.e. we look for conformal blocks
which are not uniquely defined {\it functions} of $x$ at given $\Delta_i$, $\Delta$, $c$. One immediate example is provided in the point
$\Delta_i=1/16$, $\Delta=1/4$, $c=1$ (see s.\ref{AT}):
this is the case of Ashkin-Teller model solved by Al.Zamolodchikov \cite{ZamAT}. On the
other hand, at the same point there is another conformal block, given by an elliptic integral which follows from formulas in
\cite{MMSh}. The existence of two different solutions may imply that this would provide us with needed non-perturbative
parameter. However, as we explain later in s.\ref{NZ}, this is not the case: the ambiguity is just due to different possibilities of approaching
the singularity, since the Ashkin-Teller model is located in the moduli space exactly at the Kac zero.

Another possibility to look for non-perturbative parameters could be in the cases when some of the external dimensions correspond
to a degenerate vector. If this vector is degenerate at level $k$ and one specially matches the intermediate dimension,
there is a differential equation of order $k$ in $x$ for the conformal block. This equations has generally $k$ independent solutions and, hence,
the full answer would be a linear combination of these, depending on $k-1$ arbitrary constants. However, these constants can not be
associated with non-perturbative ambiguity in the definition of the conformal block, since each of these $k$ solutions correspond to
exactly one conformal block fixed by a proper asymptotics! Moreover, the modular transformation $x\to 1-x$ transforms any of these
conformal blocks through remaining $k-1$.

In fact, the absence of non-perturbative parameter in the second case is not surprising: we already mentioned that
as a function of $x$ the 4-point
spherical conformal block is believed to be a function, ramified just at three points: $0,\ 1,\ \infty$,
with no essential singularities. Hence, the expansion series in $x$ have finite radia of convergence,
and in this sense they are somewhat different, say, from (\ref{facser}).
However, the convergence in $x$ is not uniform in the other parameters
$(\Delta_i,\Delta,c)$, and this makes the entire function of all
variables quite sophisticated, which we observe in the Ashkin-Teller case.

\section{Non-perturbative parameters. First quest: Ashkin-Teller model \label{AT}}

\subsection{Elliptic integrals}
\setcounter{equation}{0}

Note that (\ref{DFint}) is an
integral representation of entire conformal block ${\cal B}(x)$,
not of individual coefficients $J_k=B_k$ of its $x$-expansion.
A natural idea could be to use such representation as an obvious
candidate for a non-perturbative definition.

Of course, this idea raises a number of interesting questions --
especially, about the analytical continuation in $N$ and associated
non-perturbative dependencies on intermediate dimensions, like $\Delta$.
This would open a way to study non-perturbative modular kernels \cite{Zam,PT,GMM,Iorgov,Nem2}
(which perturbatively are just Fourier transforms \cite{Galpert}).

However, before going deeper in that direction,
it makes sense to look at this approach in a less
controversial situation: at natural (positive integer) values of $N_1$ and $N_2$.
It deserves beginning from the simplest case of $c=1$, i.e. $b=1$. In this case there is a solution of
(\ref{Nvialpha}) at $N_1=1$, $N_2=0$.

Even after that there is a further simplification: for a special choice
of external dimensions, when the integral becomes elliptic.

For instance, at $\alpha_i=-1/4$ the integral (\ref{DFint}) turns into just an ordinary elliptic integral
\be
K(x)=\int\frac{dz}{y(z)}
\ee
with $y^2(z)=z(1-z)(z-x)$, which is elementary to analyze, both perturbatively and
non-perturbatively.
\be
\int_0\frac{dz}{y(z)} = \sum c_k^2x^k = 1 +\frac{1}{4}x + \frac{9}{64}x^2+ \ldots
\ee
and
\be
{\cal B}^{ell}(x) = (1-x)^{1/8}K(x) =
1+\frac{1}{8}x+\frac{7}{128}x^2+\frac{33}{1024}x^3+\frac{713}{32768}x^4
+\frac{4165}{262144}x^5+ \\
+\frac{51205}{4194304}x^6
+\frac{326255}{33554432}x^7+\frac{17078585}{2147483648}x^8+\frac{114071265}{17179869184}x^9
+ \ldots
\label{ellexpan}
\ee
The MAPLE command to generate this formula is
\begin{verbatim}
series((1-x)^(1/8)*EllipticK(sqrt(x))/Pi*2,x,10);
\end{verbatim}

For non-perturbative analysis of integrals
the best method is via the Picard-Fuchs equations. The Picard-Fuchs equation for
\be
\Pi = \oint_C \frac{dz}{y(z)} = \oint_C \frac{dz}{\sqrt{z(1-z)(z-x)}}
\ee
along {\it any closed} contour $C$ is
\be
\left(x(1-x)\frac{\p^2}{\p x^2} + (1-2x)\frac{\p}{\p x}+\frac{1}{4}\right)\Pi = 0
\ee
There are two solutions: one is $K(x)$ having the asymptotics $1$ at small $x$, the other one, $K'(x)$ has the asymptotics $\log x$
\cite{GR}. We are definitely interested in the first case.

\subsection{Zamolodchikov's formula}

In \cite{Zamell} Al.Zamolodchikov suggested a wonderful formula
for the non-perturbative conformal block at the
slice where
$\Delta_i =\frac{1}{16},\ \ c=1$:
\be
{\cal B}^{Zam}_\Delta(x) = {\cal B}_\Delta\left(\Delta_i=\frac{1}{16},c=1\, \Big|\,x\right) =
(1-x)^{-1/8} \frac{(16q/x)^\Delta}{\theta_{00}(q)}
\ee
where relation between $x$, considered as a ramification point,  and elliptic
parameter $q=e^{i\pi\tau}$ is given by
\be
x = \frac{\theta_{10}^4}{\theta_{00}^4} = 16q\cdot
  \frac{(1+q^2+q^6+\ldots)^4}{(1+2q+2q^4+\ldots)^4} =
  16q - 128q^2 + \ldots
\label{xq}
\ee
and the theta-constants
\be
\theta_{00}(q) = \theta_{00}(0|\tau) = 1+ 2\sum_{n=1}^\infty q^{n^2}
= \prod_{m=1}^{\infty} (1- q^{2m})(1+ q^{2m-1})^2 \\
\theta_{01}(q) = \theta_{01}(0|\tau) = 1+ 2\sum_{n=1}^\infty (-)^nq^{n^2}
= \prod_{m=1}^{\infty} (1- q^{2m})(1 - q^{2m-1})^2    \\
\theta_{10}(q) = \theta_{10}(0|\tau) = 2q^{1/4}\sum_{n=0} q^{n(n+1)}
= 2q^{1/4} \prod_{m=1}^{\infty} (1- q^{2m})(1 + q^{2m})^2 \\
\theta'_{11}(q) =  \theta_{00}\theta_{01}\theta_{10}
= 2q^{1/4} \prod_{m=1}^\infty (1-q^{2m})^3 = \eta^3(q)
\ee
Equation (\ref{xq}) relates the modular transformations of $x$,
generated by $x\longrightarrow 1-x$ and $x\longrightarrow -1/x+1$
to the modular transformations of theta constants generated by
$\theta_{\epsilon,\delta}(0|-1/\tau) = (-i)^{\epsilon\delta}\sqrt{-i\tau} \theta_{\delta,\epsilon}(0|\tau)$
and
$\theta_{\epsilon,\delta}(0|\tau+1) = e^{\pi i/4\epsilon}\theta_{\epsilon,\delta+1-\epsilon}(0|\tau)$, where the characteristics
of the $\theta$-functions $\epsilon$ and $\delta$ are understood as taken by modulo 2.
The crucial role in this relation is played by the Riemann identity
$\theta_{00}^4(0|\tau) =  \theta_{01}^4(0|\tau)+ \theta_{10}^4(0|\tau)$.

At  the Ashkin-Teller point $\mu_{AT}$, where additionally $\Delta=\frac{1}{4}$,
\be
(1-x)^{1/8}{\cal B}^{Zam}_{1/4} = \frac{(16q/x)^{1/4}}{\theta_{00}} =
\frac{2q^{1/4}}{\theta_{10}} = \frac{1}{1+q^2+q^6+\ldots}
\ee
and perturbative expansion near in powers of $x$ is
\be
B_{1/4}^{Zam}(x) = \left(1+\frac{1}{8}x+ \frac{9}{128}x^2+\ldots\right)
\left(1-\frac{1}{256}x^2+\ldots\right)
 = 1+\frac{1}{8}x+\frac{17}{256}x^2+\frac{93}{2048}x^3+\frac{2269}{65536}x^4+\\
 + \frac{14705}{524288}x^5+\frac{198109}{8388608}x^6
+\frac{1370655}{67108864}x^7+\frac{77366631}{4294967296}x^8
+\frac{554104463}{34359738368}x^9 + \ldots
\label{Zamexpan}
\ee
The MAPLE command to generate this formula is
\begin{verbatim}
q:=x->EllipticNome(x);
A:=x->(1-x)^(-1/8)*(16*q(sqrt(x))/x)^(1/4)/JacobiTheta3(0,q(sqrt(x)));
series(A(x),x,10);
\end{verbatim}

\subsection{Intersection of physical and elliptic slices}

\subsubsection{The problem}

Now we can compare the two expressions (\ref{ellexpan}) and (\ref{Zamexpan}).
Both correspond to the Ashkin-Teller point
\be
\mu_{AT}: \ \ \ \Delta_i=\frac{1}{16}, \ \ \ \Delta = \frac{1}{4}, \ \ \ \ c=1
\label{ATpoint}
\ee
in the moduli space ${\cal M}$, both are expansions of the well defined
and well known functions of $x$, but they are different
already in the second terms of the $x$-expansion!

What does this mean?
How could we get two different expressions for $B_2$ at the given point?

The reason is that $\mu_{AT}$ actually lies on the divisor of $B_2$.
In order to obtain the second coefficient of the $x$-expansion from (\ref{B2})
one needs to resolve the singularity $\frac{0}{0}$, and the resolution is ambiguous.
Of course, this is a usual situation for a function of many variables
(even rational, like $B_k$), still this causes additional ambiguities in the conformal block: it is
not uniquely defined by the conformal dimensions at the Kac zeroes.

Our immediate goal in this situation is to study the vicinity
of $\mu_{AT}$ and see how the two expressions
(\ref{ellexpan}) and (\ref{Zamexpan}) emerge from a single (\ref{B2}).
At the next step we should look at other coefficients $B_k$ and at other
interesting points on the divisor.

\subsubsection{$B_2$ in the vicinity of $\mu_{AT}$}

Thus, we substitute
\be
\Delta_i = \frac{1}{16} + \epsilon\delta_i,
\ \ \ \ \ \ \Delta = \frac{1}{4} + \epsilon\delta, \ \ \ \ \
c = 1 + \epsilon\sigma
\ee
into expression (\ref{B1}), (\ref{B2}), $\ldots$ for the coefficients of
the $x$-expansion of conformal block and look at what happens at small values of $\epsilon$.

For $B_1$ we get nothing interesting:
\be
B_1^{AT} = \frac{1}{8} + O(\epsilon)
\ee
and this is exactly what is needed in {\it both} (\ref{ellexpan}) and (\ref{Zamexpan})

However, for $B_2$ the situation gets far more interesting:
\be
B_2^{AT} = \frac{\frac{25}{256}\sigma\epsilon + O(\epsilon^2)}
{\frac{3}{2}\sigma\epsilon + O(\epsilon^2)} \ \stackrel{?}{=} \
\frac{25}{384} + O(\epsilon)
\ee
This is still another rational number {\it different} from those in both
(\ref{ellexpan}) and (\ref{Zamexpan}).

However, let us still look at the next order in $\epsilon$:
\be
B_2^{AT} = \frac{\frac{25}{256}\sigma\epsilon + \rho\epsilon^2 + O(\epsilon^3)}
{\frac{3}{2}\sigma\epsilon + (16\delta^2+8\sigma\delta)\epsilon^2+ O(\epsilon^3)}
\ee
where
\be
\rho = \frac{1}{16}\Big(17\delta^2 -2 \delta(\delta_1+\delta_2+\delta_3+\delta_4)
+ 12(\delta_1+\delta_2)(\delta_3+\delta_4)\Big)
+ \frac{15}{32}\sigma\Big(2\delta -\delta_1+\delta_2+\delta_3-\delta_4\Big)
\ee

Now one can see what happens.
In both cases (\ref{ellexpan}) and (\ref{Zamexpan})
one approach the point $\mu_{AT}$ keeping $\sigma = 0$.
Then
\be
B_2^{AT}(c=1) = \frac{1}{256}\left(17 - 2\frac{\delta_1+\delta_2+\delta_3+\delta_4}{\delta}
+12\frac{(\delta_1+\delta_2)(\delta_3+\delta_4)}{\delta^2}\right) + O(\epsilon)
\label{B2ATquad}
\ee

Zamolodchikov's formula (\ref{Zamexpan}) refers to the physical slice,
when the theory and observables (external dimensions $\Delta_i$) are fixed:
thus, $\sigma=0$ and $\delta_1=\ldots=\delta_4=0$, and only the intermediate
dimension {\it could} vary.
In other words, only $\delta$ is {\it imagined} to be non-vanishing.
When we approach $\mu_{AT}$ from this special direction we get
\be
\lim_{\delta\rightarrow 0} B_2^{AT}\left(\Delta_i=\frac{1}{16},c=1\right) =
\frac{17}{256}
\ee
i.e. the answer is (\ref{Zamexpan}).

For the elliptic integral (\ref{ellexpan}) the situation is absolutely different.
What is fixed in this case is the number of integrations,
$N_1=1$, $N_2=0$.
According to (\ref{Nvialpha}) this implies that
we approach $\mu_{AT}$ from a very different direction, where
\be
\sigma=0,  \\
\delta+2(\delta_1+\delta_2)=0,  \\
\delta-2(\delta_3+\delta_4) = 0
\label{ellATquad}
\ee
Note that (\ref{Nvialpha}) is written in terms of $\alpha$-parameters,
not dimensions, and at $c=1$ our $\delta_i = \frac{\alpha_i-\alpha_i^{AT}}{2\epsilon\sqrt{\Delta_i}}$, hence,
the additional coefficient $2 = \frac{\sqrt{1/4}}{\sqrt{1/16}}$ in (\ref{ellATquad}).
Then (\ref{B2ATquad}) gives
\be
\lim_{\delta\rightarrow 0} B_2^{AT}\Big((\ref{ellATquad}),c=1\Big) =
\frac{1}{256}\Big(17 - \frac{12}{2^2}\Big) = \frac{7}{128}
\ee
i.e. exactly what is needed for (\ref{ellexpan}).
Note that (\ref{ellATquad}) imposes only two constraints,
but this turns to be enough to provide
an unambiguous limit in (\ref{B2ATquad}).

In fact, it is both convenient and natural to put the central charge
and the dimensions on equal footing.
If we parameterize $c=1 - 6(b-1/b)^2$, see (\ref{divialpha}), and use $b=1+\epsilon\eta$ instead of
$c = 1+\epsilon\sigma$, deviations from the AT point will be entirely of the order $\epsilon^2$
in the numerator and denominator.
In other words,
\be
\sigma = -24\eta^2\epsilon
\label{sigmasecor}
\ee
and this is the resolution of singularity that we use in the rest of this section.
In particular, in this parametrization
\be
B_2^{AT} = \frac{
\frac{17}{16}\delta^2-\frac{75}{32}\eta^2
- \frac{1}{8}\delta(\delta_1+\delta_2+\delta_3+\delta_4)
+\frac{3}{4}(\delta_1+\delta_2)(\delta_3+\delta_4)
\ + O(\epsilon)}{16\delta^2-36\eta^2  \ + O(\epsilon)}
\label{B2AT}
\ee
where we omitted the common overall factors $\epsilon^2$ in
the numerator and denominator.

\subsubsection{Other   $B_k$: universality and the germ of conformal block at $\mu_{AT}$}

If one now makes the same substitution in $B_3$, one again obtains the double zeroes
at $\mu_{AT}$ in the numerator and denominator, and
\be
B_3^{AT} = \frac{
\frac{7533}{1024}\delta^2-\frac{32805}{2048}\eta^2
- \frac{729}{512}\delta(\delta_1+\delta_2+\delta_3+\delta_4)
+\frac{2187}{256}(\delta_1+\delta_2)(\delta_3+\delta_4)
\ + O(\epsilon)}{162\delta^2-\frac{729}{2}\eta^2  \ + O(\epsilon)}
\ee
The numbers can look ugly, however they are in fact
just the same as in (\ref{B2AT}):
\be
B_3^{AT} =
-\frac{15}{512} + \frac{9}{8}\cdot B_2^{AT}\ + O(\epsilon)
\ee
This means that we do not need to perform any {\it independent} calculation
in the third terms in (\ref{ellexpan}) and (\ref{Zamexpan}):
inter-relation between these two cases is fully fixed at the level of
the second coefficient.

Indeed, for $B_4$ the same property persists for higher $B_k$:

\be
B_1=\frac{1}{8}, \ \ \ \ \ \
B_2=\frac{17\,-\,r}{256}, \ \ \ \ \ \
B_3=\frac{93-9\,r}{2^{12}}, \ \ \ \ \ \
B_4=\frac{2269-281\,r}{2^{16}},   \\
B_5=\frac{14705-2125\,r}{2^{19}} = \frac{5\cdot 17}{2^{19}}\cdot(173-25\,r), \ \ \ \ \ \
\ldots
\ee
where
\be
\,r\equiv
\frac{-3\eta^2+8\delta(\delta_1+\delta_2+\delta_3+\delta_4)
-48(\delta_1+\delta_2)(\delta_3+\delta_4)}{4\delta^2-9\eta^2}
\ee
It is natural to {\it assume} that this remains true in general:
\be
B_k=B^{Zam}_k+\frac{\,r}{3}\Big(B^{ell}_k - B^{Zam}_k\Big), \ \ \ \ k<6
\ee
However, this is actually true only for $k<6$. Indeed, at level 6 there Kac zero at $\Delta=1/4$ becomes of the fourth order:
$(n-m)^2/4=1/4$ when $n=2$, $m=1$, i.e. at level $n\cdot m=2$; when $n=3$, $m=2$, i.e. at level $n\cdot m=6$ etc.
The numerator still cancels this multiple zero at the Ashkin-Teller point, however, the ambiguity becomes a ratio
of two quartic polynomials of $\delta_i,\delta,\eta$.
For $6\leq k <12$ we have:
\be
B_k=B^{Zam}_k+\frac{\,r}{3}\Big(B^{ell}_k - B^{Zam}_k\Big) + {r_2\over 3\cdot 2^{11}}C_k, \ \ \ \ 6\leq k <12
\ee
with
\be
C_{k<6}=0, \ \  C_6 = 1, \ \ C_7=\frac{25}{8}, \ \ C_8 = \frac{1577}{800}C_7, \ \ldots
\ee
where
\be
r_2=
{P_4(\eta,\delta,\delta_i)\over (4\delta^2-9\eta^2)(4\delta^2-25\eta^2)}
\ee
and
\be
P_4(\eta,\delta,\delta_i)=-60\eta^4+\eta^2\Big(15\delta^2+86\delta
(\delta_1+\delta_2+\delta_3+\delta_4)+120(\delta_1+\delta_2)^2+120(\delta_3+\delta_4)^2
-660(\delta_1+\delta_2)(\delta_3+\delta_4)\Big)-\\
-8\delta^3(\delta_1+\delta_2+\delta_3+\delta_4)+
48\delta^2\Big(3(\delta_1+\delta_2)(\delta_3+\delta_4)-(\delta_1+\delta_2)^2-(\delta_3+\delta_4)^2\Big)
-\\-
192\delta(\delta_1+\delta_2)(\delta_3+\delta_4)(\delta_1+\delta_2+\delta_3+\delta_4)+
960(\delta_1+\delta_2)^2(\delta_3+\delta_4)^2
\ee

The moral of this story, is that
the behavior of conformal block in the vicinity of the point $\mu_{AT}$
is {\it universal}: does not depend on the order $k$ of the $x$-expansion.
This opens a possibility to suggest a  formula for
the {\it germ} of conformal block at $\mu_{AT}$ (the next $r_3$ emerge at level 12, when the Kac zero gets multiplicity 6):
\fr{
B(x)=B^{Zam}(x)+\frac{r}{3}\Big(B^{ell}(x) - B^{Zam}(x)\Big)+\\+  {r_2\over 3}\cdot
{x^6\over 2^{11}}\Big(1+{25\over 8}x+{1577\over 256}x^2+{20141\over 2048}x^3+{911193\over 65536}x^4+{9549597\over
524288}x^5+\ldots\Big)+x^{12}B_{12}+\ldots +O(\epsilon)}
Zamolodchikov's expansion corresponds to all external $\delta_i=0$, while
the elliptic locus in the vicinity of the AT point is a union of several hyperplanes,
each  defined by two conditions:
\be
\sum_{i=1}^4\delta_i= - \frac{3}{2}\,\eta,\ \ \ \ \ \delta=2(\delta_3+\delta_4)+\eta
\ee
or
\be
\sum_{i=1}^4\delta_i=  \frac{3}{2}\,\eta,\ \ \ \ \  \delta+2(\delta_3+\delta_4)=2\eta
\ee

\subsection{Chain vectors}

As an alternative to the above technique, one can perform an analysis of the singularity locus in terms of
the representation of  $B_k$ in (\ref{Bser})
via the chain-vectors (s.3.2). For instance, at the Ashkin-Teller point the leading behaviour of the chain-vectors is given by
\be
\beta_{2}= - \frac{\zeta}{\xi} +O(\xi) \\
\beta_{11}=\frac{\zeta}{\xi} +O(\xi) \\
\beta_{21}=-  \frac{\zeta}{2\xi}+O(\xi)  \\
\beta_{111}=\frac{\zeta}{2\xi} +O(\xi) \\
\beta_{3}=0
\ee
where
\be
\zeta={\delta-6\delta_1-6\delta_2\over 2(32\delta^2+3\eta)}
\ee
and $\xi$ is a distance from the singularity locus.
Thus, the chain-vectors are non-zero vectors nearby the locus, but they must have zero norm in the leading order, since their
norm (which is equal to the conformal block (\ref{Bsca})) is finite on the locus. Indeed, the singularities cancel in the
conformal block, because of degeneracy
of the Shapovalov matrix (since the singularity locus is located in zeroes of determinant of the Shapovalov matrix)
\be
\left(\sum_{|Y|=k} {\rm sing}(\beta_Y) \right)^2 = 0
\ee

\section{Perturbative conformal block in the vicinity of Kac divisor\label{NZ}}

\setcounter{equation}{0}

\subsection{Poles of $B_k$ and their nested structure}

We already discussed in s.3 and demonstrated in the manifest Ashkin-Teller example in s.4 that
\begin{itemize}
\item
if a zero appears in the Kac determinant $K_k$, it persists in all higher $K_k$ with $k\geq m$.
The Kac determinants depend only on the intermediate dimension $\Delta$ and
the central charge $c$, and in the $\alpha$-parametrization (\ref{divialpha})
the zeroes are actually at the points (\ref{Kz}).
At such points
all the coefficients $B_k$ with $k\geq m$ are singular.
\item
However, one can adjust external dimensions $\Delta_i$ so that the numerator
in $B_m$ also vanishes.
What happens is that then it also vanishes in the numerators
of all higher $B_k$ with $k\geq m$.
\end{itemize}
For example, a
Kac zero at the third level is at
\be
\alpha_{1,3}=-b
\ee
The numerator of $B_3$ at this zero is
\be
X=\frac{X_{12}X_{34}}
{24b^5(b^2-1)(3b^2-1)(4b^4-1)}
\ee
with
\be
X_{12} =
(\alpha_1-\alpha_2)
(b^2-1-b\alpha_1-b\alpha_2)(2b^2-1-b\alpha_1-b\alpha_2)
(b+\alpha_1-\alpha_2)(b-\alpha_1+\alpha_2)(1+b\alpha_1+b\alpha_2)
\ee
The numerator of $B_4$ is the same $X$, multiplied by
\be
  \frac{(2b^3+2b+\alpha_2^2b-\alpha_2b^2+\alpha_2-\alpha_1^2b+\alpha_1b^2-\alpha_1)
  (2b^3+2b+\alpha_3^2b-\alpha_3b^2+\alpha_3-\alpha_4^2b
  +\alpha_4b^2-\alpha_4)}{4b^2(b^2+1)}
\ee
This demonstrates that at the Kac zero of level $m=3$ the zero of the numerator of $B_m$
(the zero of $X$)
remains a zero of the higher $B_k$, e.g. of $B_4$.

Once again, not only the zero loci $V(K_k)$ of Kac determinants are nested,
\be
V(K_k)\subset V(K_l) \ \ \ \ {\rm or}\ \ \ \  K_l\,\vdots\, K_k
\ \ \ \ {\rm for} \ \ \ \ l>k
\ee
the same is true for the {\it numerators} of $B_k$, provided they are restricted
to $V(K_k)$:
for ${\cal V}_k = \Big\{{\rm zeroes\ of}\ B_k\Big|_{V(K_k)}\Big\}$ we have
\be
\boxed{
{\cal V}_k\subset {\cal V}_l \ \ \ \ \ {\rm for} \ \ \ k<l
}
\label{embzerolocus}
\ee

\subsection{Coefficients $B_k$ at intersection of zeroes of numerators and denominators}

At generic point of  $\ \ \bigcup_k V(K_k)\ $ the coefficients of conformal block are singular
(have poles), thus this union form a {\it singularity locus} in the moduli space
${\cal M}$ and it has codimension one.
However, there is a codimension one hypersurface within the singularity locus
(thus it has codimension two in ${\cal M}$), where the numerators are also vanishing.
And at these points we have an ambiguity of the type $0/0$, the value
of the coefficients depending on the direction from which one approaches such point in ${\cal M}$.
It is natural to name this codimension-two hypersurface {\it the ambiguity locus}. Our next goal is to describe behavior of
the entire conformal block, not just of its particular coefficients $B_k$,
at this locus.

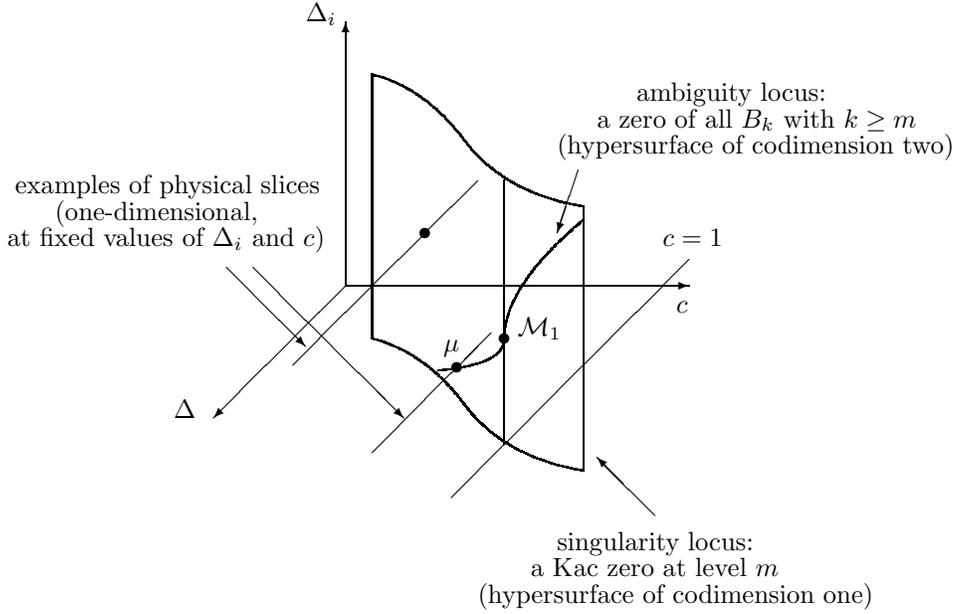
\begin{picture}(300,250)(-170,-135)
\put(0,0){\vector(1,0){130}}
\put(0,0){\vector(0,1){100}}
\put(0,0){\vector(-1,-1){50}}
\put(10,-20){\line(0,1){100}}
\put(90,-70){\line(0,1){100}}
\qbezier(10,-20)(30,-25)(45,-45)
\qbezier(90,-70)(60,-65)(45,-45)
\qbezier(10,80)(30,75)(45,55)
\qbezier(90,30)(60,35)(45,55)
\put(60,-60){\line(0,1){100}}
\qbezier(35,-32)(60,-30)(60,-20)
\qbezier(90,25)(60,0)(60,-20)
\put(130,10){\line(-1,-1){90}}
%
\put(125,-10){\mbox{$c$}}
\put(-65,-50){\mbox{$\Delta$}}
\put(-15,100){\mbox{$\Delta_i$}}
\put(80,-100){\mbox{singularity locus:}}
\put(70,-110){\mbox{a Kac zero at level $m$}}
\put(50,-120){\mbox{(hypersurface of codimension one)}}
\put(120,15){\mbox{$c=1$}}
\put(110,70){\mbox{ambiguity locus:}}
\put(95,60){\mbox{a zero of all $B_k$ with $k\geq m$}}
\put(80,50){\mbox{(hypersurface of codimension two)}}
\put(117,-87){\vector(-1,1){22}}
\put(88,44){\vector(-1,-3){8}}
\put(-20,-30){\line(1,1){70}}
\put(30,20){\circle*{4}}
\put(-125,35){\mbox{examples of physical slices}}
\put(-110,25){\mbox{(one-dimensional,}}
\put(-128,15){\mbox{at fixed values of $\Delta_i$ and $c$)}}
\put(-45,7){\vector(1,-1){30}}
\put(10,-63){\line(1,1){45}}
\put(42,-31){\circle*{4}}
\put(-35,7){\vector(1,-1){57}}
\put(37,-25){\mbox{$\mu$}}
\put(60,-20){\circle*{4}}
\put(65,-19){\mbox{${\cal M}_1$}}
\end{picture}

Of interest for us in this section will be the points $\mu\in {\cal A}\subset {\cal M}$,
lying at the ambiguity locus ${\cal A}$.
The thing is that most of interesting well known examples of conformal models,
including Ashkin-Teller and minimal models are of this type.
As we already saw in discussion of the vicinity of the point
$\mu_{AT}\in {\cal M}_1\subset{\cal A}\subset{\cal M}$,
all the coefficients $B_k$ can behave similarly near this locus,
thus their common properties are inherited by the entire
conformal block.

\bigskip

As before, at the vicinity of a point $\mu\in {\cal A}$, $\mu = \{\Delta_i^{(0)},\Delta^{(0)},c^{(0)}\}$
we put $\Delta_i = \Delta^{(0)}_i+\epsilon\delta_i$,
$\Delta=\Delta^{(0)}+\epsilon\delta$, $c=c^{(0)}+\epsilon\sigma$ or $b=b^{(0)}+\epsilon\eta$
with a $6$-vector $\xi_I=(\delta,\delta_1,\ldots,\delta_4,\eta)$,
and it turns out that
\be
B_k = b^{(0)}_k + d_k\cdot \frac{u_I\xi_I + O(\xi^2)}{v_I\xi_I+O(\xi^2)}
\ee
if the zero is simple,
\be
B_k = b^{(0)}_k +d_k\cdot \frac{U_{IJ}\xi_I\xi_J+O(\xi^3)}{V_{IJ}\xi_I\xi_J+O(\xi^3)}
\ee
if it is double zero etc.
It often happens (as in the previous section) that
\be
\boxed{
{\rm linear/quadratic\ forms}\ u,v\  {\rm and} \ U,V\
{\rm are\ {\it independent}\ of}\ k
}
\label{univers}
\ee
Note that the modified parametrization (\ref{sigmasecor})
is essential only if $c^{(0)}=1$, otherwise, $\sigma$ and $\eta$ are related
linearly without any additional damping factor $\epsilon$.

Thus at the critical point we have both an ambiguity (the limit depends on the choice
of direction to approach the point)
and a universality, expressed by (\ref{embzerolocus}) and (\ref{univers}).

In the rest of this section we consider a few more examples of
of (\ref{embzerolocus}) and (\ref{univers}).

\subsection{Phenomenon 1: nested structure of zeroes. $c=1$ example}

At $c=1$ all zeroes of the Kac determinants are doubled ($\Delta_0=\alpha^2$ is itself a
full square):
\be
K_2= 4 \Delta (4 \Delta - 1)^2, \\
K_3=72 \Delta  (4 \Delta - 1)^2(\Delta - 1)^2,\\
K_4=   2304\Delta^2 (4\Delta - 1)^2( \Delta - 1)^2 (4 \Delta - 9)^2,  \\
\ldots
\ee
The numerators of the coefficients $B_k$ at $c=1$ are nothing special. For instance,
if one takes all the external dimensions equal to each other, $\Delta_1=\ldots=\Delta_4=\Delta_e$,
the numerators of $B_2$ and $B_3$ are respectively
\be
{NB}_2=\Delta\Big(\Delta-6\Delta^2+9\Delta^3+8\Delta^4+16\Delta_e^2\Delta-
8\Delta_e\Delta+8\Delta_e\Delta^2+8\Delta_e^2\Big)
\ee
and
\be
{NB}_3=3\Delta(\Delta+2)(\Delta-1)^2\Big(8\Delta^4+19\Delta^3+24\Delta_e\Delta^2
-11\Delta^2+48\Delta_e^2\Delta-24\Delta_e\Delta+2\Delta+24\Delta_e^2\Big)
\ee
Things, however, change considerably if one looks at these numerators at the Kac zeroes.
For example, at $\Delta=\alpha_{2,1}^2=\Delta_{2,1}(c=1)=\Delta_{1,2}(c=1)=\frac{1}{4}$,
\be
{NB}_2= \frac{3}{256}
\Big(1-8(\Delta_1+\Delta_2)+16(\Delta_1^2+\Delta_2^2)-32\Delta_1\Delta_2\Big)
\Big(1-8(\Delta_3+\Delta_4)+16(\Delta_3^2+\Delta_4^2)-32\Delta_3\Delta_4\Big)
\equiv \frac{3}{256}\,nb_2,\\
{NB}_3 = -\frac{27}{16384}\Big(9+4(\Delta_1-\Delta_2)\Big) \Big(9+4(\Delta_3-\Delta_4)\Big)\cdot nb_2,
 \\
{NB}_4 \sim nb_2,  \\
\ldots  \\
{NB}_k \sim nb_2\ \ \ \ {\rm for}\ k\geq 2
\ee
This is how (\ref{embzerolocus}) is realized in this case: any zero of ${NB}_2$
remains zero of higher ${NB}_k$.

Let us now look at the vicinity of a double zero of this ${NB}_2$. In this case $\Delta_1=\Delta_2\pm \sqrt{\Delta_2}+1/4$
$\Delta_4=\Delta_3\pm \sqrt{\Delta_3}+1/4$ and one has (choosing for the sake of definiteness both signs plus):
\be
B_2={\sqrt{\Delta_2\Delta_3}(4\sqrt{\Delta_2\Delta_3}-2\sqrt{\Delta_2}-2\sqrt{\Delta_3}+3)\over 4}+
{\sqrt{\Delta_2\Delta_3}(2\sqrt{\Delta_2}+1)(2\sqrt{\Delta_3}+1)\over 4}\ r
\ee
\be
B_3={\sqrt{\Delta_2\Delta_3}(20\Delta_2\Delta_3+6\Delta_2\sqrt{\Delta_3}+6\Delta_3\sqrt{\Delta_2}+
4\Delta_2+4\Delta_3+27\sqrt{\Delta_2\Delta_3}-6\sqrt{\Delta_2}-6\sqrt{\Delta_3}+8)\over 18}
\\+{(\sqrt{\Delta_2}-2)(\sqrt{\Delta_3}-2)\over 18}{\sqrt{\Delta_2\Delta_3}(2\sqrt{\Delta_2}+1)(2\sqrt{\Delta_3}+1)\over 4}\ r
\\
\ldots
\ee
where
\be
r\equiv {3\eta^2-4\delta(\hat\delta_1+\hat\delta_2+
\hat\delta_3+\hat\delta_4)-6(\hat\delta_1\hat\delta_3+\hat\delta_2\hat\delta_4)+12\hat\delta_1\hat\delta_4+3\hat\delta_2\hat\delta_3
\over 4\delta^2-9\eta^2}
\ee
and we introduced rescaled quantities: $\hat\delta_1=\delta_1/(2\sqrt{\Delta_2}+1)$,  $\hat\delta_2=\delta_2/\sqrt{\Delta_2}$,
 $\hat\delta_3=\delta_3/\sqrt{\Delta_3}$,  $\hat\delta_4=\delta_4/(2\sqrt{\Delta_3}+1)$.
This is how (\ref{univers}) works in the $c=1$ case.

Similarly, substituting $\Delta=\alpha_{3,1}^2=\Delta_{3,1}(c=1)=\Delta_{1,3}(c=1)=1$, one gets:
\be
{NB}_3 = -18(\Delta_1-\Delta_2)\Big(1-2(\Delta_1+\Delta_2)+(\Delta_1-\Delta_2)^2\Big)
(\Delta_3-\Delta_4)\Big(1-2(\Delta_3+\Delta_4)+(\Delta_3-\Delta_4)^2\Big)
 \\
{NB}_4 = 100(4-\Delta_1+\Delta_2)(4+\Delta_3-\Delta_4)\cdot {NB}_3,  \\
\ldots  \\
{NB}_k \sim NB_3\ \ \ \ {\rm for}\ k\geq 3
\ee
Thus, one observes that any zero of ${NB}_3$
remains zero of higher ${NB}_k$.

One can easily check that this remains true for other Kac zeroes.

\subsection{Phenomenon 2: nested structure is not always enhanced at multiple (irregular) poles. $c=7/10$ and $c=1/2$ examples}

At $c=7/10$ there is a simple pole at $\Delta=3/2$ on the third level, i.e. at $B_3$, while in $B_4$ it becomes the double pole.
This is because at $b=\sqrt{5}/2$ there is
an additional degeneracy: $\Delta_{1,3}=\Delta_{4,1}=\frac{3}{2}$.

When all $\Delta_i=3/2$, there is only a first-order zero in the numerators
of both $B_3$ and $B_4$, so that $B_4$ remains infinite, while $B_3$ is just ambiguous, which
is not like the cases we considered earlier. For instance, in the Ashkin-Teller case a non-zero multiplicity of the pole
immediately resulted into the non-zero multiplicity of the corresponding zero of the conformal block numerator. This means
that the in $c=7/10$ case the structure constants $C_{3/2,3/2}^{3/2}$ should vanish (while in the Ashkin-Teller case there is no need
for this).
In fact, this imposes restrictions on the rational conformal theories, since at rational values of the central charge
there always emerge poles of higher multiplicities in the conformal block.

In $c=7/10$ theory the problem emerges in the symmetric point, when all the external dimensions are equal to each other.
However, a similar phenomenon takes place already in a simpler $c=1/2$ theory though in a non-symmetric point. We now present this case
in a little more detail.

At the central charge $c=1/2$ the Kac determinants are
\be
K_2= 2 \Delta_0 (16 \Delta_0 - 1)(2\Delta_0-1), \\
K_3=6 \Delta_0  (16 \Delta_0 - 1)(2 \Delta_0 - 1)^2  (3 \Delta_0 - 5),\\
K_4=   6\Delta_0 (16\Delta_0 - 1)^2(2 \Delta_0 - 1)^2 (16 \Delta_0 - 21)
(3 \Delta_0 - 5)  (2 \Delta_0 - 7),  \\
\ldots
\ee
The double zeroes in these formulas occur due to coincidence of the dimensions at $c=1/2$:
$\Delta_{1,3}=\Delta_{2,1}=\frac{1}{2}$ and $\Delta_{2,2}=\Delta_{1,2}=\frac{1}{16}$.
However, these accidental enhanced zeroes do not produce extra poles in $B_3$ and $B_4$:
for arbitrary values of the four external dimensions the numerator of $B_3\sim (2\Delta_0-1)$,
and $B_4\sim (16\Delta_0-1)(2\Delta_0-1)$, and this guarantees that the poles remain simple.

Nevertheless, the situation turns out to be not that simple. The remaining simple zero still needs to be compensated in the numerator,
and now this is not universal. Indeed, let us consider the conformal block of $c=1/2$ theory
with first two external dimensions parameterized as $\Delta_1={x^2-1\over 48}+{1\over 3}+{x\over 6}$ and
$\Delta_2={x^2-1\over 48}$, which guarantees that the zero is simple in the denominator of $B_2$. However, in $B_3$ this does not
provide an extra zero in addition to the factor $(2\Delta_0-1)$ which emerges independently of the external dimensions. 
This breaks the nested structure
of (\ref{embzerolocus}), and one has to impose an additional restriction to avoid infinities.

\subsection{Phenomenon 3: universality. $c=1/2$ example}

The leading behaviour is the vicinity of the singularity is
\be
B_3=
{(x+5)(x-1)(x+2)(\Delta_3-\Delta_4)(2\Delta_3^2-4\Delta_3\Delta_4+2\Delta_4^2-3\Delta_3-3\Delta_4+1)\over 3^4 2^3(3\sigma-7\delta)}
{1\over\epsilon}+O(\epsilon^0)
\ee
Hence, one can either choose a particular $x$, or specially match $\Delta_3$ and $\Delta_4$.

In the first case, one can choose, for instance, $x=-2$. Then, the nested structure is restored: this condition is enough to cancel
poles in $B_3,\ B_4,\ \ldots$. However, the universality (\ref{univers}) is broken down similarly to the Ashkin-Teller case
in the previous section: the conformal block looks like
\be
B_1={2\Delta_3-\Delta_4+1\over 4}\\
B_2= 
{1\over 7\cdot 2^5}\Big[36(\Delta_3^2+\Delta_4^2-\Delta_3\Delta_4)+88\Delta_3-80\Delta_4+31\Big]+
{1\over 7\cdot 2^5}(12\Delta_3^2-24\Delta_3\Delta_4+12\Delta_4^2-8\Delta_3-8\Delta_4+1)\ r\\
B_{24\ge k\ge 3}=B^{(0)}_k+B^{(1)}_kr+B^{(2)}_kr_2\\
\ldots
\ee
since the pole at $\Delta=1/2$ becomes triple at level 25: $\Delta_{(5,5)}=1/2$ at $c=1/2$. Here
\be
r\equiv {\sigma-28\delta_1-28\delta_2\over 7\delta+2\sigma}\\
r_2\equiv {\delta_1-\delta_2\over 3\sigma-7\delta}
\ee

In the second case, for matching $\Delta_3$ and $\Delta_4$ one can use
the same parametrization:  $\Delta_4={y^2-1\over 48}+{1\over 3}+{y\over 6}$ and
$\Delta_3={y^2-1\over 48}$ so that 
\be
B_3=
{5\over 3^7 2^7}{(x+5)(x-1)(x+2)(y+5)(y-1)(y+2)\over 3\sigma-7\delta}
{1\over\epsilon}+O(\epsilon^0)
\ee
and one suffices to choose $y=1$, $y=-2$ or $y=-5$. However, in this case of two restricted dimensions and one matched to
cancel the pole in the denominator, i.e. in the case of only one-parametric subspace in the 4-dimensional space 
$(\Delta_1,\Delta_2,\Delta_3,\Delta_4$), one can observe a new phenomenon.

\subsection{Phenomenon 4: additional universality of minimal models. Ising model example}

In the case of the four external dimensions, parameterized by two variables $(x,y)$ as above (so that appropriate zero occurs
in the numerator), the conformal block is equal 
at, say, $x=-2$ to
\be 
B_1={1-y\over 2}\\
B_2={(y-1)(y-7)\over 3\cdot 2^6}\\
B_3=-{5\over 3^4 2^8}(y-1)(y-7)(y-13)-{5\over 3^4 2^6}(y-1)(y+2)(y+5)r\\
B_4= {1\over 7\cdot 3^5 2^{13}}(y-1)(145y^3-4575y^2+58143y-211825)+{5\over 3^5 2^8}(y-19)(y+2)(y+5)(y-1)r\\
\ldots
\ee
with
\be
r={\delta_1-\delta_2\over 3\sigma-7\delta}
\ee
One can see that with such a choice of the external dimensions, the coefficient $B_2$ is unambiguous, 
while the ambiguity goes away in higher $B_k$ 
as soon as one chooses $y=1,-2,-5$, since the coefficients in front of $r$ are cancelled in such a case. Note now that
the external dimensions $(\Delta_3,\Delta_4)$ is $(0,1/2)$ at $y=1$, $(1/16,1/16)$ at $y=-2$ and $(1/2,0)$ at $y=-5$. These
dimensions are exactly the ones of the Ising model: the model with the central charge $c=1/2$,
and the three primary operators:
with dimensions $0$, $1/2$ and $1/16$.

Thus, in the Ising model we encounter a new phenomenon: {\bf the coefficient in front of $r$ vanishes}, i.e.
\be
{\bf dependence\ on\ direction\ of\ approach\ to\ the\ Ising\ point\ disappears}
\ee
despite the Ising point lies on the ambiguity locus. 

In fact, this remains the case for other minimal models (see an example of $c=1$ case in Appendix B),
and thus (if there are no other examples of this kind)
can serve as still another definition of minimal models,
entirely at the level of perturbative conformal blocks.

Let us see how it works with various 4-point conformal blocks
in the vicinity of the Ising point.

\paragraph{$\Delta_e^{(0)}=\frac{1}{16}$, $\Delta^{(0)} = \frac{1}{2}$}
\be
B_2 = \frac{\frac{9}{32}(2\sigma+7\delta)\epsilon+O(\epsilon^2)}{2(2\sigma+7\delta)\epsilon+
O(\epsilon^2)} = \frac{9}{64} + O(\epsilon),  \\
B_3 = \frac{\frac{75}{128}(2\sigma+7\delta\Big)\Big(3\sigma-7\delta\Big)\epsilon^2 + O(\epsilon^3)}
{6(2\sigma+7\delta)(3\sigma-7\delta)\epsilon^2+O(\epsilon^3)}
= \frac{25}{256} + O(\epsilon),  \\
B_4 = \frac{\frac{502047}{2048}(2\sigma+7\delta)(3\sigma-7\delta)\epsilon^2+O(\epsilon^3)}
{12\cdot 273\cdot(2\sigma+7\delta)(3\sigma-7\delta)\epsilon^2+O(\epsilon^3)}
= \frac{613}{8192} + O(\epsilon),  \\
\ldots
\ee

\bigskip

\paragraph{$\Delta_e^{(0)}=\frac{1}{16}$, $\Delta^{(0)} = 0$}
\be
B_2 = \frac{\frac{1}{32}\delta\epsilon + O(\epsilon^2)}{2\delta\epsilon + O(\epsilon^2)} =
\frac{1}{64} +O(\epsilon),  \\
B_3= \frac{\frac{15}{32}\delta\epsilon+O(\epsilon^2)}{30\delta\epsilon+O(\epsilon^2)} =
\frac{1}{64} +O(\epsilon),  \\
B_4 = \frac{-\frac{257985}{4096}\delta\epsilon+O(\epsilon^2)}{-4410\delta\epsilon+O(\epsilon^2)}=
\frac{117}{8192} + O(\epsilon)
\ee
again there is no $r$-dependence.

See more details about the Ising model in Appendix A.

\subsection{Summary}

In this section we originated a detailed examination of Kac zeroes,
where most of conventionally studied conformal models are located,
and where the standard near-divisor ambiguity arises,
preventing definition of conformal blocks,
both perturbative and non-perturbative,
as the well-defined limit from non-singular expressions.
We saw in the previous section 4 that this ambiguity can explain
the apparent difference between available non-perturbative
conformal blocks at the Ashkin-Teller point,
and thus we believe that understanding of the near-divisor
structure will be important for further development of
non-perturbative CFT.
As we explained, already at the first glance,
this structure is quite interesting.
Namely, we described four non-trivial phenomena
specific for conformal blocks and emphasizing that they
are far from exhibiting a {\it generic} behavior near
the singularity: quite the opposite, their behavior
is adjusted in a very special way, which should be better
studied and interpreted.

\bigskip

So, when we look at the Kac zero at level $l$,
the corresponding singularity appears first in the coefficient $B_l$,
and to avoid singularity one should adjust the external dimensions $\Delta_i$
to make the numerator of $NB_l$ vanishing as well.
Then the pole continues to be present in all the higher coefficients
$B_k$ with $k\geq l$. But:
\begin{itemize}
\item {\bf Phenomenon 1.} There is a nested structure in conformal blocks: the zero of the {\it numerator}
$NB_l$ at this pole is also present in all higher numerators $NB_k$
with $k\geq l$. This means that when the Kac zero is simple,
it is enough to adjust external dimensions only once, in the first
relevant $B_l$, and then the
entire conformal block is non-singular: the zero of the first relevant
numerator at the simple Kac zero is inherited by all other numerators.
\end{itemize}
However, for many interesting choices of internal (intermediate) dimension
$\Delta$, the zero of Kac determinant is {\it not} simple,
there are "accidental" coincidences of different zeroes in the numerators
of $B_k$ with $k\geq l$ for all conformal theories with rational central charges. Then
\begin{itemize}
\item {\bf Phenomenon 2.} The nested structure gets broken, in the sense that the
zero in the numerator of conformal block remains simple.
Thus, for accidentally degenerate Kac zeroes we encounter "naked singularities"
like in {\it generic} function on the moduli space.
\end{itemize}
However, there is a notable exception from this pessimistic picture:
\begin{itemize}
\item {\bf Phenomenon 3.} The nested structure is typically restored, as soon as one additionally adjusts 
the cancellation of the multiple pole in the first relevant $B_l$. In this case,
the intermediate $\Delta$ is such that the Kac zeroes are not simple,
but one and the same choice of external $\Delta_i$ provides all the numerators
with the zeroes of exactly the right order to eliminate the singularity. Then, the ambiguity in the conformal block
is piece-wise universal: if the simple Kac zero emerges at some $l_1$, it becomes a double zero at some $l_2$ etc (as soon
as there emerges an "accidental" double pole, sooner or later there emerge all higher multiplicities),
all $B_{k<l_1}$ are unambiguous, $B_{l_1\le k<l_2}$ are universal linear functions of one parameter describing approach to the
singularity locus, $B_{l_2\le k<l_3}$ are universal functions which are linear combinations of two parameters etc.
\end{itemize}
More than that:
\begin{itemize}
\item {\bf Phenomenon 4.}  At the minimal model points the near-divisor ambiguity
{\it disappears}: the limit does {\it not} depend on the direction on the moduli
space, from which we approach the minimal model points.
This is probably the most spectacular manifestation of how special the minimal models
really are from the point of view of conformal block properties,
and relation of this property to many others (like the finiteness
of block quantity, needed for the conformal bootstrap at these points)
still remains to be understood. 
\end{itemize}

It is not clear if these four phenomena provide
an exhaustive description of peculiarities
of the conformal block behavior at the Kac divisor,
even in the simplest 4-point spherical case.
Very interesting should be extension of this study
to more points and higher genera.
And, of course, the crucial question is the
implication for non-perturbative corrections.
All this remains to be thoroughly investigated.

\section{Null-vectors, equations and hidden parameters}

\setcounter{equation}{0}

In the previous sections we studied ambiguities that appear at the singularity locus of the conformal block and clarified their origin
and peculiar properties.
Here we consider the specific conformal blocks when  some of the external
dimensions correspond to a degenerate vector, since in this case one can deal with the conformal block not as a series but a
space of solutions to a differential equation. Hence, one can check if it is possible to find some non-perturbative hidden parameters,
i.e. if a conformal block can be presented as a linear combination of solutions with some arbitrary coefficients.

We start the simplest example of the vector degenerate at the second level. This vector is of the form
$\tilde V = \Big(\xi L_{-1}^2 -  L_{-2}\Big)V_\Delta$
and there are two non-trivial conditions: $ L_1 \tilde V = 0$
and $ L_2 \tilde V = 0$. They imply respectively that
\be
\xi = \frac{3}{2(2\Delta + 1)}
\ee
and
\be
8\Delta + c = 12\xi\Delta
\ee
or, together
\be
\Delta = \frac{5-c \pm \sqrt{(c-1)(c-25)}}{16}
\ee
Parameterizing the central charge and dimension as in (\ref{divialpha}),
we obtain four solutions:
\be
\boxed{
\left\{ \begin{array}{c} \alpha = \frac{1}{2b} \\ \xi = b^2 \end{array}\right.,}\ \ \ \
\left\{ \begin{array}{c} \alpha = -\frac{b}{2}  \\ \xi = \frac{1}{b^2} \end{array}\right.,\ \ \ \
\left\{ \begin{array}{c} \alpha = \frac{3b}{2}-\frac{1}{b}  \\ \xi = b^2 \end{array}\right.,\ \ \ \
\left\{ \begin{array}{c} \alpha = b - \frac{3}{2b}  \\ \xi = \frac{1}{b^2} \end{array}\right.
\ee
In what follows we work with the first of these four solutions  (boxed),
so that the original highest weight primary $V_{1/2b}$
of degenerate Verma module has dimension
\be
\label{ddg}
\Delta_{1/2b} = -\frac{1}{2} + \frac{3}{4b^2}
\ee
The conformal Ward identities imply that 4-point correlators $\Psi_4(x,\bar x)$
with insertion of this degenerate primary at point $x$ satisfy peculiar
differential equations, see \cite{CFT}:
 \be \label{hg1}
\left\{ b^2x(x-1)\p_x^2 + (2x-1)\p_x + \Delta_{1/2b} +
\frac{\Delta_1}{x} - \frac{\Delta_3}{x-1} - \Delta_4\right\}
\Psi_4(x) =0
\ee
where we suppressed the dependence on $\bar x$.
In the free field realization of conformal field theory this constraint
is imposed almost automatically, see \cite{MMMsrf}, and this is also
easily seen from the Dotsenko-Fateev $\beta$-ensemble
representation of the corresponding conformal blocks, \cite{MMMsrf}.

Conjugation with a factor
$x^\alpha(1-x)^\beta$ with specially adjusted $\alpha$ and $\beta$
converts (\ref{hg1}) into an ordinary hypergeometric equation with the
solution
\be \label{hg2}
\Psi_4(x) =
x^{\alpha_1/b}(1-x)^{\alpha_3/b}F(A,B;C;x)
\\
A = {1\over 2b^2}+{\alpha_1\over b}+{\alpha_2\over b}-{\alpha_3\over b}
\\
B = {1\over b}\sum_{i=1}^3\alpha_i + 2\Delta_{1/2b},\ \ \ C={1\over b^2} + {2\alpha_1\over b}
\ee
{\bf Equations (\ref{hg1}), (\ref{hg2}) are consistent with generic formulas (\ref{B1})-(\ref{B2}) {\it only}
if the dimensions $\Delta_1$ and $\Delta$
are related (in parametrization (\ref{divialpha})) by the fusion rule}
\be
\label{fusion}
\alpha = \alpha_1 \pm {1\over 2b}
\ee
where two choices of the sign correspond to the two linearly independent solutions of
(\ref{hg1}) and in the case of the sign ``minus'' in (\ref{fusion}) one has to choose
in (\ref{hg2}) instead of $F(A,B;C;x)$ the other
solution to the hypergeometric equation so that $\alpha_1\to b-1/b-\alpha_1$ in $\Psi_4(x)$ in (\ref{hg2}).

One can easily check directly that the conformal block from the r.h.s. of
(\ref{ccb4dg})
\be
\label{B4ptdg}
{\cal B}_{\Delta_\alpha}^{(1,{1/2b};34)}(x) = x^{\Delta_\alpha-\Delta_1-\Delta_{1/2b}}
\left(1 + {(\Delta_\alpha+\Delta_{1/2b}-\Delta_1)(\Delta_\alpha+\Delta_3-\Delta_4)
\over 2\Delta_\alpha}x + \ldots\right)
\stackrel{(\ref{fusion})}{=}\ \Psi_4(x)
\ee
which solves (\ref{hg1}). Formula (\ref{CBC}) now acquires the form
\be
\label{ccb4dg}
\left< V_1(0)V_{1/2b}(x)V_3(1)V_4(\infty)\right> =
\sum_{\Delta,\bar\Delta} C_{1,{1/2b}}^{\Delta,\bar\Delta} C_{34}^{\Delta,\bar\Delta}
{\cal B}_\Delta^{(1,{1/2b};34)}(x)\bar {\cal B}_{\bar\Delta}^{(1,{1/2b};34)}(\bar x)=\\
= \sum_{\alpha = \alpha_1 \pm {1\over 2b}\atop\bar\alpha = \bar\alpha_1 \pm {1\over 2b}} 
C_{1,{1/2b}}^{\Delta_\alpha,\Delta_{\bar\alpha}} C_{34}^{\Delta_\alpha,\Delta_{\bar\alpha}}
{\cal B}_\Delta^{(1,{1/2b};34)}(x)\bar {\cal B}_{\bar\Delta}^{(1,{1/2b};34)}(\bar x)
\ee
since {\bf {\em only} for the choice (\ref{fusion}) the structure constant $C_{1,{1/2b}}^{\Delta_\alpha}$
is non-vanishing} \cite{CFT}. Here we obtained this fact indirectly by solving the
equation for the correlator. One can derive this fact straightforwardly
using the $\beta$-ensemble representation for the conformal blocks \cite{MMMsrf}.

Thus, two solutions of the equation for the degenerate conformal block describes two {\it different} conformal blocks, the only
two with non-zero structure constants. One can easily see also (\ref{fusion}) from their asymptotics: since the conformal
block behaves at small $x$ like $x^{\Delta-\Delta_1-\Delta_2}$, one gets (\ref{fusion}) from the asymptotic behaviours
$\Psi_4^{(1)}(x)\sim x^{\alpha_1/b}$ and $\Psi_4^{(2)}(x)\sim x^{1-1/b^2-\alpha_1/b}$.

This system of two conformal blocks is self-consistent: it survives the modular transformation (\ref{dmt}):
$x\to 1-x$, $\Delta_1\leftrightarrow \Delta_3$. Indeed (see \cite[(9.131(1,2))]{GR}),
\be
\Psi_4^{(1)}(\Delta_3,\Delta_2,\Delta_1,\Delta_4;1-x)=\displaystyle{{\Gamma\Big({2\alpha_3\over b}+{1\over b^2}\Big)
\Gamma\Big({1\over b^2}+{2\alpha_1\over b}-1\Big)\over
\Gamma\Big({3\over 2b^2}-1+{\alpha_1\over b}+{\alpha_3\over b}+{\alpha_4\over b}\Big)
\Gamma\Big({1\over 2b^2}+{\alpha_1\over b}+{\alpha_3\over b}-{\alpha_4\over b}\Big)}}\Psi_4^{(1)}(\Delta_1,\Delta_2,\Delta_3,\Delta_4;x)
+\\+
\displaystyle{{\Gamma\Big({2\alpha_3\over b}+{1\over b^2}\Big)
\Gamma\Big(1-{1\over b^2}-{2\alpha_1\over b}\Big)\over
\Gamma\Big(1-{1\over 2b^2}+{\alpha_3\over b}-{\alpha_1\over b}-{\alpha_4\over b}\Big)
\Gamma\Big({1\over 2b^2}+{\alpha_3\over b}+{\alpha_4\over b}-{\alpha_1\over b}\Big)}}\Psi_4^{(2)}(\Delta_1,\Delta_2,\Delta_3,\Delta_4;x)\\
\Psi_4^{(2)}(\Delta_3,\Delta_2,\Delta_1,\Delta_4;1-x)=\displaystyle{{\Gamma\Big(2-{2\alpha_3\over b}-{1\over b^2}\Big)
\Gamma\Big({1\over b^2}-1+{2\alpha_1\over b}\Big)\over
\Gamma\Big({1\over 2b^2}+{\alpha_1\over b}+{\alpha_4\over b}-{\alpha_3\over b}\Big)
\Gamma\Big(1-{1\over 2b^2}+{\alpha_1\over b}-{\alpha_3\over b}-{\alpha_4\over b}\Big)}}\Psi_4^{(1)}(\Delta_1,\Delta_2,\Delta_3,\Delta_4;x)
+\\+
\displaystyle{{\Gamma\Big(2-{2\alpha_3\over b}-{1\over b^2}\Big)
\Gamma\Big(1-{1\over b^2}-{2\alpha_1\over b}\Big)\over
\Gamma\Big(2-{3\over 2b^2}-{\alpha_1\over b}-{\alpha_3\over b}-{\alpha_4\over b}\Big)
\Gamma\Big(1-{1\over 2b^2}+{\alpha_1\over b}+{\alpha_3\over b}-{\alpha_4\over b}\Big)}}\Psi_4^{(2)}(\Delta_1,\Delta_2,\Delta_3,\Delta_4;x)
\ee
i.e. the modular kernel in this case is the $2\times 2$ matrix
\be
\left(
\begin{array}{cc}
\displaystyle{{\Gamma\Big({2\alpha_3\over b}+{1\over b^2}\Big)
\Gamma\Big({1\over b^2}+{2\alpha_1\over b}-1\Big)\over
\Gamma\Big({3\over 2b^2}-1+{\alpha_1\over b}+{\alpha_3\over b}+{\alpha_4\over b}\Big)
\Gamma\Big({1\over 2b^2}+{\alpha_1\over b}+{\alpha_3\over b}-{\alpha_4\over b}\Big)}}&
\displaystyle{{\Gamma\Big({2\alpha_3\over b}+{1\over b^2}\Big)
\Gamma\Big(1-{1\over b^2}-{2\alpha_1\over b}\Big)\over
\Gamma\Big(1-{1\over 2b^2}+{\alpha_3\over b}-{\alpha_1\over b}-{\alpha_4\over b}\Big)
\Gamma\Big({1\over 2b^2}+{\alpha_3\over b}+{\alpha_4\over b}-{\alpha_1\over b}\Big)}}\\
\displaystyle{{\Gamma\Big(2-{2\alpha_3\over b}-{1\over b^2}\Big)
\Gamma\Big({1\over b^2}-1+{2\alpha_1\over b}\Big)\over
\Gamma\Big({1\over 2b^2}+{\alpha_1\over b}+{\alpha_4\over b}-{\alpha_3\over b}\Big)
\Gamma\Big(1-{1\over 2b^2}+{\alpha_1\over b}-{\alpha_3\over b}-{\alpha_4\over b}\Big)}}&
\displaystyle{{\Gamma\Big(2-{2\alpha_3\over b}-{1\over b^2}\Big)
\Gamma\Big(1-{1\over b^2}-{2\alpha_1\over b}\Big)\over
\Gamma\Big(2-{3\over 2b^2}-{\alpha_1\over b}-{\alpha_3\over b}-{\alpha_4\over b}\Big)
\Gamma\Big(1-{1\over 2b^2}+{\alpha_1\over b}+{\alpha_3\over b}-{\alpha_4\over b}\Big)}}
\end{array}
\right)
\ee
We demonstrate how all this works in the minimal models in examples of $c=1/2$ (the Ising model) and $c=1$ in Appendices A and B
respectively.

Similarly one can deal with conformal blocks degenerate at higher levels. In these case the order of the corresponding differential
equation is higher, but again is equal exactly to the number of conformal blocks with non-zero structure constants.
The modular matrix also accordingly increases its size, see examples in \cite{CFT}. Hence, no hidden parameters are emerge in this way.
In fact, this is not surprising, since the differential equations are w.r.t. the variable $x$, and, as we already stressed, the
$x$-behaviour of the conformal block is not expected to depend on non-perturbative (hidden) parameters.

\section{Conclusion}
\setcounter{equation}{0}

The main goal of this paper is to urge the study
of the non-perturbative conformal block as a function of
{\it all} its variables:  coordinates, external and
internal dimensions and the central charge.
The first question to ask here is if
there are some {\it extra} parameters, besides already
enumerated, on which the
non-perturbative quantity usually depends, which are
not seen at the perturbative level like the
theta-angle in instanton calculus.
In conformal block story, the main ambiguity 
is the overall normalization, which can be
a {\it function} of dimensions and central charge (thus
in fact can contain infinitely many extra parameters).
The lack of control over such normalization factors is the
main current problem in relating different efficient
non-perturbative approaches, say, to constructing the modular kernel: the $SL_q(2)$
method of Ponsot-Teshner \cite{PT},
the matrix model approach of \cite{GMM}
uncovering the Stokes (wall crossing) phenomena and
relating cluster variables to check-exponents of \cite{AMMcheck},
and the Painleve equation method of \cite{Iorgov}.
In the present paper we did not attack this problem of
$\Delta$-dependence directly: instead we tried to look
for extra non-perturbative parameters, considering
the $x$-dependence of the simplest (4-point spherical) conformal block.
We looked mostly at two obvious places:
at discrepancy between the explicitly known
non-perturbative Zamolodchikov and elliptic-integral
answers at the Ashkin-Teller point (they are different functions of $x$),
and at the higher order differential equation in $x$, which
conformal block satisfies when one of the vertex
operators is degenerate (e.g. in the Ising and other minimal
models), both cases could seem to imply the
existence of extra parameters.
As we explained, this is, however, not the case.
The discrepancy at the Ashkin-Teller point (and in many similar cases)
is in fact just the ordinary ambiguity at the singularity divisor
for a function of many variables (dimensions
and central charge), and no extra parameters are
present.
The case with many, rather than one, solutions to
a higher order differential equation is resolved {\it not}
by introduction of extra variables, but by the fact that
$B(x)$ actually does {\it not} satisfy such equations
when just one external dimension is fixed:
in fact, the equation is true for the conformal block only when the internal dimension $\Delta$
is fixed as well \cite{MMMsrf} (this is a very important
feature of the conformal block, which is often overlooked
or underestimated).
In result, the extra solutions
are in fact describing not an ambiguity in the function $B_\Delta(x)$,
but the {\it other} conformal blocks $B_{\Delta'}(x)$ with the {\it other}
allowed values of internal dimension $\Delta'$:
there exactly as many of them as the degeneration level of
vertex operator and the order of the differential equation.

Thus we found {\it no} evidence for extra non-perturbative
parameters in the $x$-dependence of conformal block.
This  could seem obvious from the very beginning:
as we already mentioned, there is a belief that the 4-point spherical conformal block
is actually a Belyi function (i.e. has only non-essential singularities at
three points $x=0,1,\infty$, and ramification orders are integer for rational conformal models).
This, in turn, can be attributed either to the fact that $2d$ CFT
is actually a free field theory \cite{CFT} so that there is actually
no interaction and no reason for {\it real} non-perturbative effects
to exist, or to another fact: that it is conformal, and then no non-trivial
dependence is allowed for a function of the single dimensional parameter $x$.

We believe that our simple consideration sheds some new light on the
problem of non-perturbative conformal blocks and can help to attract
new attention to this extremely interesting   problem.
Conformal blocks are the crucially important special function of the
string era, and they should be thoroughly investigated and understood.

\section*{Acknowledgements}

Our work is partly supported by the grant
NSh-1500.2014.2 (A.M.'s),
by RFBR  13-02-00457 (A.Mir.), 13-02-00478 (A.Mor.),
by joint grants 13-02-91371-ST (A.M.'s), 14-01-92691-Ind (A.M.'s),
by the Brazil National Counsel of Scientific and
Technological Development (A.Mor.).
The research of H.I. is supported in part by the Grant-in-Ad for Scientific Research
(23540316) from the Ministry of Education,
Science and Culture, Japan. Support from JSPS/RFBR bilateral collaboration "Synthesis of integrabilities
arising from gauge-string duality"
(FY2010-2011: 12-02-92108-Yaf-a) is gratefully appreciated.

\section*{Appendix A. Ising model}
\def\theequation{A\arabic{equation}}
\setcounter{equation}{0}

The critical behaviour of the Ising model is described by the central charge $c=1/2$, and we choose $b=\sqrt{3}/2$.
There three primary fields with the dimensions: $\Delta_I=0$ (i.e. $\alpha_I=0$ or $-\sqrt{3}/6$), $\Delta_\psi=1/2$
(i.e. $\alpha_\psi=
\sqrt{3}/3$ or $-\sqrt{3}/2$), $\Delta_\sigma=1/16$
(i.e. $\alpha_\sigma=\sqrt{3}/12$ or $-\sqrt{3}/4$).
The first field is degenerate
at the first level, the second and the third ones are degenerate at the second level.
One can calculate the
conformal blocks in different cases.
For instance, consider ${\cal B}(\Delta_1,\Delta_2,\Delta_3,\Delta_4;\Delta;c;x)$
and suppose that the intermediate dimension $\Delta=\Delta_\psi=1/2$.
It leads to poles at all levels higher than one.
The condition of canceling this pole by matching the dimensions $\Delta_1$ and $\Delta_2$
requires them be either
$1/2$ and 0 or $\Delta_1=\Delta_2=1/16$
(if one restricts himself with the spectrum of three fields above).
Hence, the correlators of fields $<\psi\psi I>$ and $<\psi\sigma\sigma>$ are non-zero.
Similarly considering
the conformal block with the intermediate dimension $\Delta=\Delta_\sigma=1/16$,
one finds the non-zero correlator
$<\psi\psi I>$.
This fixes the operator product expansion (OPE) of fields.

Now consider the conformal block with the field $\psi(x)$ at point $x$
(remind that, in our notation, this corresponds to
$\Delta_2$, while $\Delta_1$, $\Delta_3$ and $\Delta_4$ corresponds
to the fields at points 0, 1 and $\infty$
respectively). This field is degenerate at the second level and
the conformal block satisfies a second order
differential equation provided $\alpha=\alpha_1\pm 1/2b$ (see s.6).
This condition can be
satisfied only for the pairs ($\alpha,\alpha_1$) either
($\alpha_\sigma$,$\alpha_\sigma$) or ($\alpha_\psi,\alpha_I$)
which again fixes the OPE.
Consider the correlator $<\sigma(0)\psi(x)\sigma(1)\psi(\infty)>$.
It is described
by the values $\alpha_1=-\sqrt{3}/4$, $\alpha_2=\sqrt{3}/3$,
$\alpha_3=-\sqrt{3}/4$, $\alpha_4=-\sqrt{3}/2$.
The corresponding conformal block ${\cal B}(x)$ satisfies the differential equation (\ref{hg1})
\be\label{deq}
\left[{3\over 4}x(1-x)\partial^2_x+(2x-1)\partial_x+
{1\over 16}\left({1\over x}+{1\over 1-x}\right)\right]{\cal B}(x)=0
\ee
This equation is hypergeometric and has two solutions:
\be
{\cal B}^{(1)}(x)={1-2x\over\sqrt{x(1-x)}}\\
{\cal B}^{(2)}(x)=[x(1-x)]^{1/6}F(1/3,2;5/3;x)
\ee
where $F(a,b;c;x)$ is the hypergeometric function.
The first solution corresponds to the behaviour at small $x$
\be\label{se}
{\cal B}^{(1)}(x)={1\over\sqrt{x}}\left(1-{3\over 2}x-{5\over 8}x^2
-{7\over 16}x^3-{45\over 128}x^4+\ldots\right)
\ee
of the conformal block. The multiplier $1/\sqrt{x}$ comes from the usual pre-factor $x^{\Delta-\Delta_1-\Delta_2}$
of the conformal block and implies (which we already
established from the OPE earlier) that $\Delta=\Delta_\sigma=1/16$.
The second solution ought to describe an
intermediate field with dimension $\Delta=35/48$.
This field is absent in the spectrum which means that the
corresponding structure constant is zero.

The expansion (\ref{se}) of the conformal block should be
compared with its generic expansion (\ref{cbe}).
One can check that they coincide independently on the way  the singularity is resolved,
i.e. if one considers a vicinity of the point $\Delta_1=1/16$, $\Delta_2=1/2$, $\Delta_3=16$,
$\Delta_4=1/2$, $\Delta=1/16$ , the leading order in $\epsilon$ does not depend on the way of approaching the singularity
at all (see s.5.6):
\be
(\ref{cbe})=1-{3\over 2}x-{5\over 8}x^2-{7\over 16}x^3-{45\over 128}x^4+\ldots
\ee

Note that ${\cal B}^{(1)}(x)$ is consistently invariant with respect
 to the duality transformation $x\to 1-x$.
Similarly invariant is the differential equation (\ref{deq}),
though the second solution is not: it transforms
through itself and the first solution \cite[9.131(2)]{GR}:
\be
F(1/3,2;5/3;x)=F(1/3,2;5/3;1-x)+{\Gamma (2/3)^3\over\sqrt{3}\pi}{1\over[x(1-x]^{1/6}}{\cal B}^{(1)}(x)
\ee
This means that the duality matrix is triangle.

Now consider another possible 4-point correlator in this theory:
$<\sigma\sigma\sigma\sigma>$.
One can similarly
right down the differential equation
\be
\left[{4\over 3}x(1-x)\partial^2_x+(2x-1)\partial_x
+{1\over 16}\left({1\over x}+{1\over 1-x}\right)\right]{\cal B}(x)=0
\ee
This equation has two solutions:
\be
{\cal B}^{(1)}(x)={\sqrt{\sqrt{x}+1}+\sqrt{\sqrt{x}-1}\over [x(1-x]^{1/8}}\\
{\cal B}^{(2)}(x)={\sqrt{\sqrt{x}+1}-\omega\sqrt{\sqrt{x}-1}\over [x(1-x]^{1/8}}
\ee
where $\omega\equiv \exp(\pm \pi i/2)$ is a square root of -1
(plus or minus depends on the chosen branch of
$\sqrt{x}$). The first solution has the small-$x$ expansion
\be
{\cal B}^{(1)}(x)\sim {1\over x^{1/8}}
\left(1+ {1\over 64}x^2
+{1\over 64}x^3+{117\over 8192}x^4+{53\over 4096}x^5+\ldots\right)
\ee
and corresponds to the intermediate field $I$ with dimension $\Delta=\Delta_I=0$.
The second solution has
the small-$x$ expansion
\be
{\cal B}^{(2)}(x)\sim {\sqrt{x}\over x^{1/8}}\left(1+{1\over 4}x
+{9\over 64}x^2+{25\over 256}x^3+{613\over 8192}x^4\right)
\ee
These two expansions as before reproduce the correct result (\ref{se})
independently on the way of resolving
the singularity.

Since, in this case, the both solutions correspond to the "physical" conformal blocks,
this is not surprising
that the duality transformation acts as a matrix on these two solutions:
\be
{\cal B}^{(1)}(1-x)={1\over\sqrt{2}}\left({\cal B}^{(1)}(x)+{\omega\over 2}{\cal B}^{(2)}(x)\right)\\
{\cal B}^{(2)}(1-x)=\sqrt{2}\left({\cal B}^{(1)}(x)-i{\omega\over 2}{\cal B}^{(2)}(x)\right)
\ee
i.e. the duality matrix is
\be
S=\sqrt{2}\left(
\begin{array}{lr}
\displaystyle{1}&\displaystyle{{\omega\over 2}}\\
\\
\displaystyle{2}&\displaystyle{-i\omega}
\end{array}
\right)
\ee

The last non-trivial correlator in the Ising model is $<\psi\psi\psi\psi>$.
It is described by the differential equation
\be
\left[{3\over 4}x(1-x)\partial^2_x+(2x-1)\partial_x+{1\over 2}
\left({1\over x}+{1\over 1-x}\right)\right]{\cal B}(x)=0
\ee
Again, only one of the two solutions of this equation is relevant to the Ising model, it is
\be
{\cal B}(x)={1-x+x^2\over x(1-x)}
\ee
This solution is modular invariant and its small-$x$ expansion gives the conformal block
(\ref{se}) independently on the way of resolving the singularity:
\be
{\cal B}(x)\sim {1\over x}\left(1+x^2+x^3+x^4+\ldots\right)
\ee
where the common factor $1/x$ implies that this conformal block describes
the intermediate field of dimension
$\Delta=\Delta_I=0$ as it should be.

\section*{Appendix B. $c=1$, $\Delta_e=\frac{1}{4}$}

\def\theequation{B\arabic{equation}}
\setcounter{equation}{0}

Let us consider the minimal model with $c=1$, it can be obtained from the series of minimal models $(m,m+1)$ in the limit $m\to\infty$
\cite[App.B]{DF}, and one easily construct the conformal block of four fields with $\Delta_{(1,2)}=1/4$, since they are degenerate at the 
second level. Solving equation (\ref{hg1}) with $b=1$ gives two solutions which correspond to the conformal block with internal 
dimensions $\Delta=0$:
\be
{\cal B}_{\Delta = 0}\left(\Delta_i=\frac{1}{4},\,c=1\,\Big|\,x\right) =
{1\over\sqrt{x}}\Big(1+\frac{x^2}{8}+\frac{x^3}{8}+\frac{15x^4}{128} + \ldots \Big)=
\frac{1}{2\sqrt{x}}\left(\sqrt{1-x}+\frac{1}{\sqrt{1-x}}\right)
\ee
and $\Delta=1$
\be
{\cal B}_{\Delta=1}\left(\Delta_i=\frac{1}{4},\,c=1\,\Big|\,x\right) =
\sqrt{x}\Big(1+\frac{x}{2}+\frac{3x^2}{8}+\frac{5x^3}{16}+\frac{35x^4}{128} + \ldots \Big)=
\frac{\sqrt{x}}{\sqrt{1-x}}
\ee

The modular transformation $x \longrightarrow 1-x$ acts on the doublet
\be
\left(
\begin{array}{c}
{1\over 2}\sqrt{{1-x\over x}}+{1\over 2}\sqrt{{1\over x(1-x)}}\\
\sqrt{{x\over 1-x}}
\end{array}
\right)
\ee
by the matrix
\be
\left(
\begin{array}{cc}
1/2 & 3/4 \\
1 & -1/2
\end{array}
\right)
\ee
From bilinear combination of these one can construct a modular invariant by adjusting the coefficient:
\be
\frac{1}{|x|}\left|{\cal B}_0\right|^2 + \frac{3}{4}|x|\cdot\left|{\cal B}_1\right|^2 =
\frac{\left|1-\frac{x}{2}\right|^2 + \frac{3}{4}\cdot|x|^2}{|x|\cdot|1-x|}
\sim \frac{|x|}{|1-x|}+\frac{|1-x|}{|x|}+\frac{1}{|x(1-x)|}
\ee
This answer coincides with \cite{DF}.

\end{document}